\documentclass[12pt,mathserif]{article}
\usepackage{lmodern}
\usepackage{amssymb,amsmath}
\usepackage{ifxetex,ifluatex}
\usepackage{fixltx2e} % provides \textsubscript
\ifnum 0\ifxetex 1\fi\ifluatex 1\fi=0 % if pdftex
  \usepackage[T1]{fontenc}
  \usepackage[utf8]{inputenc}
\else % if luatex or xelatex
  \ifxetex
    \usepackage{mathspec}
  \else
    \usepackage{fontspec}
  \fi
  \defaultfontfeatures{Ligatures=TeX,Scale=MatchLowercase}
\fi
\IfFileExists{upquote.sty}{\usepackage{upquote}}{}
\IfFileExists{microtype.sty}{%
\usepackage{microtype}
\UseMicrotypeSet[protrusion]{basicmath} % disable protrusion for tt fonts
}{}
\usepackage[margin=1in]{geometry}
\usepackage{hyperref}
\hypersetup{unicode=true,
            pdfborder={0 0 0},
            breaklinks=true}
\usepackage{graphicx,grffile}
\makeatletter
\def\maxwidth{\ifdim\Gin@nat@width>\linewidth\linewidth\else\Gin@nat@width\fi}
\def\maxheight{\ifdim\Gin@nat@height>\textheight\textheight\else\Gin@nat@height\fi}
\makeatother
\setkeys{Gin}{width=\maxwidth,height=\maxheight,keepaspectratio}
\IfFileExists{parskip.sty}{%
\usepackage{parskip}
}{% else
\setlength{\parindent}{0pt}
\setlength{\parskip}{6pt plus 2pt minus 1pt}
}
\providecommand{\tightlist}{%
  \setlength{\itemsep}{0pt}\setlength{\parskip}{0pt}}
\let\rmarkdownfootnote\footnote%
\def\footnote{\protect\rmarkdownfootnote}

\usepackage{titling}

  \title{}
    \pretitle{\vspace{\droptitle}}
  \posttitle{}
    \author{}
    \preauthor{}\postauthor{}
    \date{}
    \predate{}\postdate{}
  
\usepackage{hyperref}
\usepackage{graphics}
\usepackage{longtable}
\usepackage[absolute,overlay]{textpos}

\pdfpageattr {/Group << /S /Transparency /I true /CS /DeviceRGB>>}       

\usepackage{amssymb,amsmath, amsbsy, amsthm, graphicx,setspace,rotating} % necessary for some symbols, and apparently text in equations.

\usepackage{etex}
\usepackage[T1]{fontenc}
\usepackage{inputenc}
\usepackage {standalone}
\usepackage{times}
\usepackage[notocbib]{apacite}
\usepackage{graphicx} % to incorporate pdfs
\usepackage{multirow}
\usepackage[import]{xy}
\usepackage{rotating}
\usepackage{booktabs}
\usepackage{listings}
\usepackage[english]{babel}
\usepackage[sc]{mathpazo}
\usepackage{pgf}
\usepackage{pdflscape} %% for inclusion of single page landscape

\newcommand{\blandscape}{\begin{landscape}}
\newcommand{\elandscape}{\end{landscape}}

\usepackage{wasysym} % Checkboxes
\usepackage[normalem]{ulem}

\usepackage[bf,tiny,compact]{titlesec}
\titleformat*{\section}{\bfseries\centering}
\titlespacing{\section}{0pc}{*0.1}{*0.1}[0pc]
\titlespacing{\subsection}{0pc}{*0.1}{*0.1}[0pc]
\titlespacing{\subsubsection}{0pc}{*0.1}{*0.1}[0pc]
\titlespacing{\paragraph}{0pc}{*0.1}{*1}[]
\allowdisplaybreaks[4]

\usepackage{tikz}
\usetikzlibrary{arrows,decorations.pathmorphing,backgrounds,positioning,fit,petri,shapes.geometric}
\tikzstyle{block}=[draw opacity=0.7,line width=1.4cm]
\usetikzlibrary{arrows}
\usetikzlibrary{positioning}
\usetikzlibrary{shadows}
\usetikzlibrary{automata}
\usetikzlibrary{backgrounds}
\usetikzlibrary{calc}
\usetikzlibrary{shapes.callouts}
\usetikzlibrary{shapes.arrows}
\tikzset{every loop/.style={min distance=3mm,looseness=2}}

\tikzstyle{ov}=[shape=rectangle,
                drop shadow,
                draw=black,
	              fill=gray!20,
                minimum height=0.6cm,
                minimum width=0.6cm]

\tikzstyle{av}=[shape=rectangle,
	      drop shadow,
                draw=black,
                fill=gray!20,
                fill=black!10,
                minimum height=0.6cm,
                minimum width=0.6cm]
\tikzstyle{lv}=[shape=circle,drop shadow, draw=black,fill=gray!20,minimum width=0.7cm]

\usepackage{float} % forces placement of Figures w/ [H] option
\usepackage{fancyhdr}
\begin{document}

% Basic Greek Lower Hat
\newcommand{\alphahat}{\hat{\alpha}}
\newcommand{\betahat}{\hat{\beta}}
\newcommand{\gammahat}{\hat{\gamma}}
\newcommand{\deltahat}{\hat{\delta}}
\newcommand{\epsilonhat}{\hat{\epsilon}}
\newcommand{\varepsilonhat}{\hat{\varepsilon}}
\newcommand{\zetahat}{\hat{\zeta}}
\newcommand{\etahat}{\hat{\eta}}
\newcommand{\thetahat}{\hat{\theta}}
\newcommand{\varthetahat}{\hat{\vartheta}}
\newcommand{\iotahat}{\hat{\iota}}
\newcommand{\kappahat}{\hat{\kappa}}
\newcommand{\lambdahat}{\hat{\lambda}}
\newcommand{\muhat}{\hat{\mu}}
\newcommand{\nuhat}{\hat{\nu}}
\newcommand{\xihat}{\hat{\xi}}
\newcommand{\pihat}{\hat{\pi}}
\newcommand{\varpihat}{\hat{\varpi}}
\newcommand{\rhohat}{\hat{\rho}}
\newcommand{\varrhohat}{\hat{\varrho}}
\newcommand{\sigmahat}{\hat{\sigma}}
\newcommand{\varsigmahat}{\hat{\varsigma}}
\newcommand{\tauhat}{\hat{\tau}}
\newcommand{\upsilonhat}{\hat{\upsilon}}
\newcommand{\phihat}{\hat{\phi}}
\newcommand{\varphihat}{\hat{\varphi}}
\newcommand{\chihat}{\hat{\chi}}
\newcommand{\psihat}{\hat{\psi}}
\newcommand{\omegahat}{\hat{\omega}}

% Basic Greek Lower Tilde
\newcommand{\alphatilde}{\tilde{\alpha}}
\newcommand{\betatilde}{\tilde{\beta}}
\newcommand{\gammatilde}{\tilde{\gamma}}
\newcommand{\deltatilde}{\tilde{\delta}}
\newcommand{\epsilontilde}{\tilde{\epsilon}}
\newcommand{\varepsilontilde}{\tilde{\varepsilon}}
\newcommand{\zetatilde}{\tilde{\zeta}}
\newcommand{\etatilde}{\tilde{\eta}}
\newcommand{\thetatilde}{\tilde{\theta}}
\newcommand{\varthetatilde}{\tilde{\vartheta}}
\newcommand{\iotatilde}{\tilde{\iota}}
\newcommand{\kappatilde}{\tilde{\kappa}}
\newcommand{\lambdatilde}{\tilde{\lambda}}
\newcommand{\mutilde}{\tilde{\mu}}
\newcommand{\nutilde}{\tilde{\nu}}
\newcommand{\xitilde}{\tilde{\xi}}
\newcommand{\pitilde}{\tilde{\pi}}
\newcommand{\varpitilde}{\tilde{\varpi}}
\newcommand{\rhotilde}{\tilde{\rho}}
\newcommand{\varrhotilde}{\tilde{\varrho}}
\newcommand{\sigmatilde}{\tilde{\sigma}}
\newcommand{\varsigmatilde}{\tilde{\varsigma}}
\newcommand{\tautilde}{\tilde{\tau}}
\newcommand{\upsilontilde}{\tilde{\upsilon}}
\newcommand{\phitilde}{\tilde{\phi}}
\newcommand{\varphitilde}{\tilde{\varphi}}
\newcommand{\chitilde}{\tilde{\chi}}
\newcommand{\psitilde}{\tilde{\psi}}
\newcommand{\omegatilde}{\tilde{\omega}}

% Basic Greek Lower Check
\newcommand{\alphacheck}{\check{\alpha}}
\newcommand{\betacheck}{\check{\beta}}
\newcommand{\gammacheck}{\check{\gamma}}
\newcommand{\deltacheck}{\check{\delta}}
\newcommand{\epsiloncheck}{\check{\epsilon}}
\newcommand{\varepsiloncheck}{\check{\varepsilon}}
\newcommand{\zetacheck}{\check{\zeta}}
\newcommand{\etacheck}{\check{\eta}}
\newcommand{\thetacheck}{\check{\theta}}
\newcommand{\varthetacheck}{\check{\vartheta}}
\newcommand{\iotacheck}{\check{\iota}}
\newcommand{\kappacheck}{\check{\kappa}}
\newcommand{\lambdacheck}{\check{\lambda}}
\newcommand{\mucheck}{\check{\mu}}
\newcommand{\nucheck}{\check{\nu}}
\newcommand{\xicheck}{\check{\xi}}
\newcommand{\picheck}{\check{\pi}}
\newcommand{\varpicheck}{\check{\varpi}}
\newcommand{\rhocheck}{\check{\rho}}
\newcommand{\varrhocheck}{\check{\varrho}}
\newcommand{\sigmacheck}{\check{\sigma}}
\newcommand{\varsigmacheck}{\check{\varsigma}}
\newcommand{\taucheck}{\check{\tau}}
\newcommand{\upsiloncheck}{\check{\upsilon}}
\newcommand{\phicheck}{\check{\phi}}
\newcommand{\varphicheck}{\check{\varphi}}
\newcommand{\chicheck}{\check{\chi}}
\newcommand{\psicheck}{\check{\psi}}
\newcommand{\omegacheck}{\check{\omega}}

% Basic Greek Lower Dot
\newcommand{\alphadot}{\dot{\alpha}}
\newcommand{\betadot}{\dot{\beta}}
\newcommand{\gammadot}{\dot{\gamma}}
\newcommand{\deltadot}{\dot{\delta}}
\newcommand{\epsilondot}{\dot{\epsilon}}
\newcommand{\varepsilondot}{\dot{\varepsilon}}
\newcommand{\zetadot}{\dot{\zeta}}
\newcommand{\etadot}{\dot{\eta}}
\newcommand{\thetadot}{\dot{\theta}}
\newcommand{\varthetadot}{\dot{\vartheta}}
\newcommand{\iotadot}{\dot{\iota}}
\newcommand{\kappadot}{\dot{\kappa}}
\newcommand{\lambdadot}{\dot{\lambda}}
\newcommand{\mudot}{\dot{\mu}}
\newcommand{\nudot}{\dot{\nu}}
\newcommand{\xidot}{\dot{\xi}}
\newcommand{\pidot}{\dot{\pi}}
\newcommand{\varpidot}{\dot{\varpi}}
\newcommand{\rhodot}{\dot{\rho}}
\newcommand{\varrhodot}{\dot{\varrho}}
\newcommand{\sigmadot}{\dot{\sigma}}
\newcommand{\varsigmadot}{\dot{\varsigma}}
\newcommand{\taudot}{\dot{\tau}}
\newcommand{\upsilondot}{\dot{\upsilon}}
\newcommand{\phidot}{\dot{\phi}}
\newcommand{\varphidot}{\dot{\varphi}}
\newcommand{\chidot}{\dot{\chi}}
\newcommand{\psidot}{\dot{\psi}}
\newcommand{\omegadot}{\dot{\omega}}

% Basic Greek Lower DDot
\newcommand{\alphaddot}{\ddot{\alpha}}
\newcommand{\betaddot}{\ddot{\beta}}
\newcommand{\gammaddot}{\ddot{\gamma}}
\newcommand{\deltaddot}{\ddot{\delta}}
\newcommand{\epsilonddot}{\ddot{\epsilon}}
\newcommand{\varepsilonddot}{\ddot{\varepsilon}}
\newcommand{\zetaddot}{\ddot{\zeta}}
\newcommand{\etaddot}{\ddot{\eta}}
\newcommand{\thetaddot}{\ddot{\theta}}
\newcommand{\varthetaddot}{\ddot{\vartheta}}
\newcommand{\iotaddot}{\ddot{\iota}}
\newcommand{\kappaddot}{\ddot{\kappa}}
\newcommand{\lambdaddot}{\ddot{\lambda}}
\newcommand{\muddot}{\ddot{\mu}}
\newcommand{\nuddot}{\ddot{\nu}}
\newcommand{\xiddot}{\ddot{\xi}}
\newcommand{\piddot}{\ddot{\pi}}
\newcommand{\varpiddot}{\ddot{\varpi}}
\newcommand{\rhoddot}{\ddot{\rho}}
\newcommand{\varrhoddot}{\ddot{\varrho}}
\newcommand{\sigmaddot}{\ddot{\sigma}}
\newcommand{\varsigmaddot}{\ddot{\varsigma}}
\newcommand{\tauddot}{\ddot{\tau}}
\newcommand{\upsilonddot}{\ddot{\upsilon}}
\newcommand{\phiddot}{\ddot{\phi}}
\newcommand{\varphiddot}{\ddot{\varphi}}
\newcommand{\chiddot}{\ddot{\chi}}
\newcommand{\psiddot}{\ddot{\psi}}
\newcommand{\omegaddot}{\ddot{\omega}}

% Basic Greek Upper Hat
\newcommand{\Gammahat}{\hat{\Gamma}}
\newcommand{\Deltahat}{\hat{\Delta}}
\newcommand{\Thetahat}{\hat{\Theta}}
\newcommand{\Lambdahat}{\hat{\Lambda}}
\newcommand{\Xihat}{\hat{\Xi}}
\newcommand{\Pihat}{\hat{\Pi}}
\newcommand{\Sigmahat}{\hat{\Sigma}}
\newcommand{\Upsilonhat}{\hat{\Upsilon}}
\newcommand{\Phihat}{\hat{\Phi}}
\newcommand{\Psihat}{\hat{\Psi}}
\newcommand{\Omegahat}{\hat{\Omega}}

% Basic Greek Upper Tilde
\newcommand{\Gammatilde}{\tilde{\Gamma}}
\newcommand{\Deltatilde}{\tilde{\Delta}}
\newcommand{\Thetatilde}{\tilde{\Theta}}
\newcommand{\Lambdatilde}{\tilde{\Lambda}}
\newcommand{\Xitilde}{\tilde{\Xi}}
\newcommand{\Pitilde}{\tilde{\Pi}}
\newcommand{\Sigmatilde}{\tilde{\Sigma}}
\newcommand{\Upsilontilde}{\tilde{\Upsilon}}
\newcommand{\Phitilde}{\tilde{\Phi}}
\newcommand{\Psitilde}{\tilde{\Psi}}
\newcommand{\Omegatilde}{\tilde{\Omega}}

% Basic Greek Upper Check
\newcommand{\Gammacheck}{\check{\Gamma}}
\newcommand{\Deltacheck}{\check{\Delta}}
\newcommand{\Thetacheck}{\check{\Theta}}
\newcommand{\Lambdacheck}{\check{\Lambda}}
\newcommand{\Xicheck}{\check{\Xi}}
\newcommand{\Picheck}{\check{\Pi}}
\newcommand{\Sigmacheck}{\check{\Sigma}}
\newcommand{\Upsiloncheck}{\check{\Upsilon}}
\newcommand{\Phicheck}{\check{\Phi}}
\newcommand{\Psicheck}{\check{\Psi}}
\newcommand{\Omegacheck}{\check{\Omega}}

% Basic Greek Upper Dot
\newcommand{\Gammadot}{\dot{\Gamma}}
\newcommand{\Deltadot}{\dot{\Delta}}
\newcommand{\Thetadot}{\dot{\Theta}}
\newcommand{\Lambdadot}{\dot{\Lambda}}
\newcommand{\Xidot}{\dot{\Xi}}
\newcommand{\Pidot}{\dot{\Pi}}
\newcommand{\Sigmadot}{\dot{\Sigma}}
\newcommand{\Upsilondot}{\dot{\Upsilon}}
\newcommand{\Phidot}{\dot{\Phi}}
\newcommand{\Psidot}{\dot{\Psi}}
\newcommand{\Omegadot}{\dot{\Omega}}

% Basic Greek Upper DDot
\newcommand{\Gammaddot}{\ddot{\Gamma}}
\newcommand{\Deltaddot}{\ddot{\Delta}}
\newcommand{\Thetaddot}{\ddot{\Theta}}
\newcommand{\Lambdaddot}{\ddot{\Lambda}}
\newcommand{\Xiddot}{\ddot{\Xi}}
\newcommand{\Piddot}{\ddot{\Pi}}
\newcommand{\Sigmaddot}{\ddot{\Sigma}}
\newcommand{\Upsilonddot}{\ddot{\Upsilon}}
\newcommand{\Phiddot}{\ddot{\Phi}}
\newcommand{\Psiddot}{\ddot{\Psi}}
\newcommand{\Omegaddot}{\ddot{\Omega}}

% Bold Greek Lower
\newcommand{\bfalpha}{\boldsymbol{\alpha}}
\newcommand{\bfbeta}{\boldsymbol{\beta}}
\newcommand{\bfgamma}{\boldsymbol{\gamma}}
\newcommand{\bfdelta}{\boldsymbol{\delta}}
\newcommand{\bfepsilon}{\boldsymbol{\epsilon}}
\newcommand{\bfvarepsilon}{\boldsymbol{\varepsilon}}
\newcommand{\bfzeta}{\boldsymbol{\zeta}}
\newcommand{\bfeta}{\boldsymbol{\eta}}
\newcommand{\bftheta}{\boldsymbol{\theta}}
\newcommand{\bfvartheta}{\boldsymbol{\vartheta}}
\newcommand{\bfiota}{\boldsymbol{\iota}}
\newcommand{\bfkappa}{\boldsymbol{\kappa}}
\newcommand{\bflambda}{\boldsymbol{\lambda}}
\newcommand{\bfmu}{\boldsymbol{\mu}}
\newcommand{\bfnu}{\boldsymbol{\nu}}
\newcommand{\bfxi}{\boldsymbol{\xi}}
\newcommand{\bfpi}{\boldsymbol{\pi}}
\newcommand{\bfvarpi}{\boldsymbol{\varpi}}
\newcommand{\bfrho}{\boldsymbol{\rho}}
\newcommand{\bfvarrho}{\boldsymbol{\varrho}}
\newcommand{\bfsigma}{\boldsymbol{\sigma}}
\newcommand{\bfvarsigma}{\boldsymbol{\varsigma}}
\newcommand{\bftau}{\boldsymbol{\tau}}
\newcommand{\bfupsilon}{\boldsymbol{\upsilon}}
\newcommand{\bfphi}{\boldsymbol{\phi}}
\newcommand{\bfvarphi}{\boldsymbol{\varphi}}
\newcommand{\bfchi}{\boldsymbol{\chi}}
\newcommand{\bfpsi}{\boldsymbol{\psi}}
\newcommand{\bfomega}{\boldsymbol{\omega}}

% Bold Greek Lower Hat
\newcommand{\bfalphahat}{\boldsymbol{\hat{\alpha}}}
\newcommand{\bfbetahat}{\boldsymbol{\hat{\beta}}}
\newcommand{\bfgammahat}{\boldsymbol{\hat{\gamma}}}
\newcommand{\bfdeltahat}{\boldsymbol{\hat{\delta}}}
\newcommand{\bfepsilonhat}{\boldsymbol{\hat{\epsilon}}}
\newcommand{\bfvarepsilonhat}{\boldsymbol{\hat{\varepsilon}}}
\newcommand{\bfzetahat}{\boldsymbol{\hat{\zeta}}}
\newcommand{\bfetahat}{\boldsymbol{\hat{\eta}}}
\newcommand{\bfthetahat}{\boldsymbol{\hat{\theta}}}
\newcommand{\bfvarthetahat}{\boldsymbol{\hat{\vartheta}}}
\newcommand{\bfiotahat}{\boldsymbol{\hat{\iota}}}
\newcommand{\bfkappahat}{\boldsymbol{\hat{\kappa}}}
\newcommand{\bflambdahat}{\boldsymbol{\hat{\lambda}}}
\newcommand{\bfmuhat}{\boldsymbol{\hat{\mu}}}
\newcommand{\bfnuhat}{\boldsymbol{\hat{\nu}}}
\newcommand{\bfxihat}{\boldsymbol{\hat{\xi}}}
\newcommand{\bfpihat}{\boldsymbol{\hat{\pi}}}
\newcommand{\bfvarpihat}{\boldsymbol{\hat{\varpi}}}
\newcommand{\bfrhohat}{\boldsymbol{\hat{\rho}}}
\newcommand{\bfvarrhohat}{\boldsymbol{\hat{\varrho}}}
\newcommand{\bfsigmahat}{\boldsymbol{\hat{\sigma}}}
\newcommand{\bfvarsigmahat}{\boldsymbol{\hat{\varsigma}}}
\newcommand{\bftauhat}{\boldsymbol{\hat{\tau}}}
\newcommand{\bfupsilonhat}{\boldsymbol{\hat{\upsilon}}}
\newcommand{\bfphihat}{\boldsymbol{\hat{\phi}}}
\newcommand{\bfvarphihat}{\boldsymbol{\hat{\varphi}}}
\newcommand{\bfchihat}{\boldsymbol{\hat{\chi}}}
\newcommand{\bfpsihat}{\boldsymbol{\hat{\psi}}}
\newcommand{\bfomegahat}{\boldsymbol{\hat{\omega}}}

% Bold Greek Lower Hat
\newcommand{\bfalphabar}{\boldsymbol{\bar{\alpha}}}
\newcommand{\bfbetabar}{\boldsymbol{\bar{\beta}}}
\newcommand{\bfgammabar}{\boldsymbol{\bar{\gamma}}}
\newcommand{\bfdeltabar}{\boldsymbol{\bar{\delta}}}
\newcommand{\bfepsilonbar}{\boldsymbol{\bar{\epsilon}}}
\newcommand{\bfvarepsilonbar}{\boldsymbol{\bar{\varepsilon}}}
\newcommand{\bfzetabar}{\boldsymbol{\bar{\zeta}}}
\newcommand{\bfetabar}{\boldsymbol{\bar{\eta}}}
\newcommand{\bfthetabar}{\boldsymbol{\bar{\theta}}}
\newcommand{\bfvarthetabar}{\boldsymbol{\bar{\vartheta}}}
\newcommand{\bfiotabar}{\boldsymbol{\bar{\iota}}}
\newcommand{\bfkappabar}{\boldsymbol{\bar{\kappa}}}
\newcommand{\bflambdabar}{\boldsymbol{\bar{\lambda}}}
\newcommand{\bfmubar}{\boldsymbol{\bar{\mu}}}
\newcommand{\bfnubar}{\boldsymbol{\bar{\nu}}}
\newcommand{\bfxibar}{\boldsymbol{\bar{\xi}}}
\newcommand{\bfpibar}{\boldsymbol{\bar{\pi}}}
\newcommand{\bfvarpibar}{\boldsymbol{\bar{\varpi}}}
\newcommand{\bfrhobar}{\boldsymbol{\bar{\rho}}}
\newcommand{\bfvarrhobar}{\boldsymbol{\bar{\varrho}}}
\newcommand{\bfsigmabar}{\boldsymbol{\bar{\sigma}}}
\newcommand{\bfvarsigmabar}{\boldsymbol{\bar{\varsigma}}}
\newcommand{\bftaubar}{\boldsymbol{\bar{\tau}}}
\newcommand{\bfupsilonbar}{\boldsymbol{\bar{\upsilon}}}
\newcommand{\bfphibar}{\boldsymbol{\bar{\phi}}}
\newcommand{\bfvarphibar}{\boldsymbol{\bar{\varphi}}}
\newcommand{\bfchibar}{\boldsymbol{\bar{\chi}}}
\newcommand{\bfpsibar}{\boldsymbol{\bar{\psi}}}
\newcommand{\bfomegabar}{\boldsymbol{\bar{\omega}}}

% Bold Greek Lower Tilde
\newcommand{\bfalphatilde}{\boldsymbol{\tilde{\alpha}}}
\newcommand{\bfbetatilde}{\boldsymbol{\tilde{\beta}}}
\newcommand{\bfgammatilde}{\boldsymbol{\tilde{\gamma}}}
\newcommand{\bfdeltatilde}{\boldsymbol{\tilde{\delta}}}
\newcommand{\bfepsilontilde}{\boldsymbol{\tilde{\epsilon}}}
\newcommand{\bfvarepsilontilde}{\boldsymbol{\tilde{\varepsilon}}}
\newcommand{\bfzetatilde}{\boldsymbol{\tilde{\zeta}}}
\newcommand{\bfetatilde}{\boldsymbol{\tilde{\eta}}}
\newcommand{\bfthetatilde}{\boldsymbol{\tilde{\theta}}}
\newcommand{\bfvarthetatilde}{\boldsymbol{\tilde{\vartheta}}}
\newcommand{\bfiotatilde}{\boldsymbol{\tilde{\iota}}}
\newcommand{\bfkappatilde}{\boldsymbol{\tilde{\kappa}}}
\newcommand{\bflambdatilde}{\boldsymbol{\tilde{\lambda}}}
\newcommand{\bfmutilde}{\boldsymbol{\tilde{\mu}}}
\newcommand{\bfnutilde}{\boldsymbol{\tilde{\nu}}}
\newcommand{\bfxitilde}{\boldsymbol{\tilde{\xi}}}
\newcommand{\bfpitilde}{\boldsymbol{\tilde{\pi}}}
\newcommand{\bfvarpitilde}{\boldsymbol{\tilde{\varpi}}}
\newcommand{\bfrhotilde}{\boldsymbol{\tilde{\rho}}}
\newcommand{\bfvarrhotilde}{\boldsymbol{\tilde{\varrho}}}
\newcommand{\bfsigmatilde}{\boldsymbol{\tilde{\sigma}}}
\newcommand{\bfvarsigmatilde}{\boldsymbol{\tilde{\varsigma}}}
\newcommand{\bftautilde}{\boldsymbol{\tilde{\tau}}}
\newcommand{\bfupsilontilde}{\boldsymbol{\tilde{\upsilon}}}
\newcommand{\bfphitilde}{\boldsymbol{\tilde{\phi}}}
\newcommand{\bfvarphitilde}{\boldsymbol{\tilde{\varphi}}}
\newcommand{\bfchitilde}{\boldsymbol{\tilde{\chi}}}
\newcommand{\bfpsitilde}{\boldsymbol{\tilde{\psi}}}
\newcommand{\bfomegatilde}{\boldsymbol{\tilde{\omega}}}

% Bold Greek Lower Check
\newcommand{\bfalphacheck}{\boldsymbol{\check{\alpha}}}
\newcommand{\bfbetacheck}{\boldsymbol{\check{\beta}}}
\newcommand{\bfgammacheck}{\boldsymbol{\check{\gamma}}}
\newcommand{\bfdeltacheck}{\boldsymbol{\check{\delta}}}
\newcommand{\bfepsiloncheck}{\boldsymbol{\check{\epsilon}}}
\newcommand{\bfvarepsiloncheck}{\boldsymbol{\check{\varepsilon}}}
\newcommand{\bfzetacheck}{\boldsymbol{\check{\zeta}}}
\newcommand{\bfetacheck}{\boldsymbol{\check{\eta}}}
\newcommand{\bfthetacheck}{\boldsymbol{\check{\theta}}}
\newcommand{\bfvarthetacheck}{\boldsymbol{\check{\vartheta}}}
\newcommand{\bfiotacheck}{\boldsymbol{\check{\iota}}}
\newcommand{\bfkappacheck}{\boldsymbol{\check{\kappa}}}
\newcommand{\bflambdacheck}{\boldsymbol{\check{\lambda}}}
\newcommand{\bfmucheck}{\boldsymbol{\check{\mu}}}
\newcommand{\bfnucheck}{\boldsymbol{\check{\nu}}}
\newcommand{\bfxicheck}{\boldsymbol{\check{\xi}}}
\newcommand{\bfpicheck}{\boldsymbol{\check{\pi}}}
\newcommand{\bfvarpicheck}{\boldsymbol{\check{\varpi}}}
\newcommand{\bfrhocheck}{\boldsymbol{\check{\rho}}}
\newcommand{\bfvarrhocheck}{\boldsymbol{\check{\varrho}}}
\newcommand{\bfsigmacheck}{\boldsymbol{\check{\sigma}}}
\newcommand{\bfvarsigmacheck}{\boldsymbol{\check{\varsigma}}}
\newcommand{\bftaucheck}{\boldsymbol{\check{\tau}}}
\newcommand{\bfupsiloncheck}{\boldsymbol{\check{\upsilon}}}
\newcommand{\bfphicheck}{\boldsymbol{\check{\phi}}}
\newcommand{\bfvarphicheck}{\boldsymbol{\check{\varphi}}}
\newcommand{\bfchicheck}{\boldsymbol{\check{\chi}}}
\newcommand{\bfpsicheck}{\boldsymbol{\check{\psi}}}
\newcommand{\bfomegacheck}{\boldsymbol{\check{\omega}}}

% Bold Greek Lower dot
\newcommand{\bfalphadot}{\boldsymbol{\dot{\alpha}}}
\newcommand{\bfbetadot}{\boldsymbol{\dot{\beta}}}
\newcommand{\bfgammadot}{\boldsymbol{\dot{\gamma}}}
\newcommand{\bfdeltadot}{\boldsymbol{\dot{\delta}}}
\newcommand{\bfepsilondot}{\boldsymbol{\dot{\epsilon}}}
\newcommand{\bfvarepsilondot}{\boldsymbol{\dot{\varepsilon}}}
\newcommand{\bfzetadot}{\boldsymbol{\dot{\zeta}}}
\newcommand{\bfetadot}{\boldsymbol{\dot{\eta}}}
\newcommand{\bfthetadot}{\boldsymbol{\dot{\theta}}}
\newcommand{\bfvarthetadot}{\boldsymbol{\dot{\vartheta}}}
\newcommand{\bfiotadot}{\boldsymbol{\dot{\iota}}}
\newcommand{\bfkappadot}{\boldsymbol{\dot{\kappa}}}
\newcommand{\bflambdadot}{\boldsymbol{\dot{\lambda}}}
\newcommand{\bfmudot}{\boldsymbol{\dot{\mu}}}
\newcommand{\bfnudot}{\boldsymbol{\dot{\nu}}}
\newcommand{\bfxidot}{\boldsymbol{\dot{\xi}}}
\newcommand{\bfpidot}{\boldsymbol{\dot{\pi}}}
\newcommand{\bfvarpidot}{\boldsymbol{\dot{\varpi}}}
\newcommand{\bfrhodot}{\boldsymbol{\dot{\rho}}}
\newcommand{\bfvarrhodot}{\boldsymbol{\dot{\varrho}}}
\newcommand{\bfsigmadot}{\boldsymbol{\dot{\sigma}}}
\newcommand{\bfvarsigmadot}{\boldsymbol{\dot{\varsigma}}}
\newcommand{\bftaudot}{\boldsymbol{\dot{\tau}}}
\newcommand{\bfupsilondot}{\boldsymbol{\dot{\upsilon}}}
\newcommand{\bfphidot}{\boldsymbol{\dot{\phi}}}
\newcommand{\bfvarphidot}{\boldsymbol{\dot{\varphi}}}
\newcommand{\bfchidot}{\boldsymbol{\dot{\chi}}}
\newcommand{\bfpsidot}{\boldsymbol{\dot{\psi}}}
\newcommand{\bfomegadot}{\boldsymbol{\dot{\omega}}}

% Bold Greek Lower ddot
\newcommand{\bfalphaddot}{\boldsymbol{\ddot{\alpha}}}
\newcommand{\bfbetaddot}{\boldsymbol{\ddot{\beta}}}
\newcommand{\bfgammaddot}{\boldsymbol{\ddot{\gamma}}}
\newcommand{\bfdeltaddot}{\boldsymbol{\ddot{\delta}}}
\newcommand{\bfepsilonddot}{\boldsymbol{\ddot{\epsilon}}}
\newcommand{\bfvarepsilonddot}{\boldsymbol{\ddot{\varepsilon}}}
\newcommand{\bfzetaddot}{\boldsymbol{\ddot{\zeta}}}
\newcommand{\bfetaddot}{\boldsymbol{\ddot{\eta}}}
\newcommand{\bfthetaddot}{\boldsymbol{\ddot{\theta}}}
\newcommand{\bfvarthetaddot}{\boldsymbol{\ddot{\vartheta}}}
\newcommand{\bfiotaddot}{\boldsymbol{\ddot{\iota}}}
\newcommand{\bfkappaddot}{\boldsymbol{\ddot{\kappa}}}
\newcommand{\bflambdaddot}{\boldsymbol{\ddot{\lambda}}}
\newcommand{\bfmuddot}{\boldsymbol{\ddot{\mu}}}
\newcommand{\bfnuddot}{\boldsymbol{\ddot{\nu}}}
\newcommand{\bfxiddot}{\boldsymbol{\ddot{\xi}}}
\newcommand{\bfpiddot}{\boldsymbol{\ddot{\pi}}}
\newcommand{\bfvarpiddot}{\boldsymbol{\ddot{\varpi}}}
\newcommand{\bfrhoddot}{\boldsymbol{\ddot{\rho}}}
\newcommand{\bfvarrhoddot}{\boldsymbol{\ddot{\varrho}}}
\newcommand{\bfsigmaddot}{\boldsymbol{\ddot{\sigma}}}
\newcommand{\bfvarsigmaddot}{\boldsymbol{\ddot{\varsigma}}}
\newcommand{\bftauddot}{\boldsymbol{\ddot{\tau}}}
\newcommand{\bfupsilonddot}{\boldsymbol{\ddot{\upsilon}}}
\newcommand{\bfphiddot}{\boldsymbol{\ddot{\phi}}}
\newcommand{\bfvarphiddot}{\boldsymbol{\ddot{\varphi}}}
\newcommand{\bfchiddot}{\boldsymbol{\ddot{\chi}}}
\newcommand{\bfpsiddot}{\boldsymbol{\ddot{\psi}}}
\newcommand{\bfomegaddot}{\boldsymbol{\ddot{\omega}}}

% Bold Greek Upper
\newcommand{\bfGamma}{\boldsymbol{\Gamma}}
\newcommand{\bfDelta}{\boldsymbol{\Delta}}
\newcommand{\bfTheta}{\boldsymbol{\Theta}}
\newcommand{\bfLambda}{\boldsymbol{\Lambda}}
\newcommand{\bfXi}{\boldsymbol{\Xi}}
\newcommand{\bfPi}{\boldsymbol{\Pi}}
\newcommand{\bfSigma}{\boldsymbol{\Sigma}}
\newcommand{\bfUpsilon}{\boldsymbol{\Upsilon}}
\newcommand{\bfPhi}{\boldsymbol{\Phi}}
\newcommand{\bfPsi}{\boldsymbol{\Psi}}
\newcommand{\bfOmega}{\boldsymbol{\Omega}}

% Bold Greek Upper Hat
\newcommand{\bfGammahat}{\boldsymbol{\hat{\Gamma}}}
\newcommand{\bfDeltahat}{\boldsymbol{\hat{\Delta}}}
\newcommand{\bfThetahat}{\boldsymbol{\hat{\Theta}}}
\newcommand{\bfLambdahat}{\boldsymbol{\hat{\Lambda}}}
\newcommand{\bfXihat}{\boldsymbol{\hat{\Xi}}}
\newcommand{\bfPihat}{\boldsymbol{\hat{\Pi}}}
\newcommand{\bfSigmahat}{\boldsymbol{\hat{\Sigma}}}
\newcommand{\bfUpsilonhat}{\boldsymbol{\hat{\Upsilon}}}
\newcommand{\bfPhihat}{\boldsymbol{\hat{\Phi}}}
\newcommand{\bfPsihat}{\boldsymbol{\hat{\Psi}}}
\newcommand{\bfOmegahat}{\boldsymbol{\hat{\Omega}}}

% Bold Greek Upper Tilde
\newcommand{\bfGammatilde}{\boldsymbol{\tilde{\Gamma}}}
\newcommand{\bfDeltatilde}{\boldsymbol{\tilde{\Delta}}}
\newcommand{\bfThetatilde}{\boldsymbol{\tilde{\Theta}}}
\newcommand{\bfLambdatilde}{\boldsymbol{\tilde{\Lambda}}}
\newcommand{\bfXitilde}{\boldsymbol{\tilde{\Xi}}}
\newcommand{\bfPitilde}{\boldsymbol{\tilde{\Pi}}}
\newcommand{\bfSigmatilde}{\boldsymbol{\tilde{\Sigma}}}
\newcommand{\bfUpsilontilde}{\boldsymbol{\tilde{\Upsilon}}}
\newcommand{\bfPhitilde}{\boldsymbol{\tilde{\Phi}}}
\newcommand{\bfPsitilde}{\boldsymbol{\tilde{\Psi}}}
\newcommand{\bfOmegatilde}{\boldsymbol{\tilde{\Omega}}}

% Bold Greek Upper Check
\newcommand{\bfGammacheck}{\boldsymbol{\check{\Gamma}}}
\newcommand{\bfDeltacheck}{\boldsymbol{\check{\Delta}}}
\newcommand{\bfThetacheck}{\boldsymbol{\check{\Theta}}}
\newcommand{\bfLambdacheck}{\boldsymbol{\check{\Lambda}}}
\newcommand{\bfXicheck}{\boldsymbol{\check{\Xi}}}
\newcommand{\bfPicheck}{\boldsymbol{\check{\Pi}}}
\newcommand{\bfSigmacheck}{\boldsymbol{\check{\Sigma}}}
\newcommand{\bfUpsiloncheck}{\boldsymbol{\check{\Upsilon}}}
\newcommand{\bfPhicheck}{\boldsymbol{\check{\Phi}}}
\newcommand{\bfPsicheck}{\boldsymbol{\check{\Psi}}}
\newcommand{\bfOmegacheck}{\boldsymbol{\check{\Omega}}}

% Bold Greek Upper Dot
\newcommand{\bfGammadot}{\boldsymbol{\dot{\Gamma}}}
\newcommand{\bfDeltadot}{\boldsymbol{\dot{\Delta}}}
\newcommand{\bfThetadot}{\boldsymbol{\dot{\Theta}}}
\newcommand{\bfLambdadot}{\boldsymbol{\dot{\Lambda}}}
\newcommand{\bfXidot}{\boldsymbol{\dot{\Xi}}}
\newcommand{\bfPidot}{\boldsymbol{\dot{\Pi}}}
\newcommand{\bfSigmadot}{\boldsymbol{\dot{\Sigma}}}
\newcommand{\bfUpsilondot}{\boldsymbol{\dot{\Upsilon}}}
\newcommand{\bfPhidot}{\boldsymbol{\dot{\Phi}}}
\newcommand{\bfPsidot}{\boldsymbol{\dot{\Psi}}}
\newcommand{\bfOmegadot}{\boldsymbol{\dot{\Omega}}}

% Bold Greek Upper DDot
\newcommand{\bfGammaddot}{\boldsymbol{\ddot{\Gamma}}}
\newcommand{\bfDeltaddot}{\boldsymbol{\ddot{\Delta}}}
\newcommand{\bfThetaddot}{\boldsymbol{\ddot{\Theta}}}
\newcommand{\bfLambdaddot}{\boldsymbol{\ddot{\Lambda}}}
\newcommand{\bfXiddot}{\boldsymbol{\ddot{\Xi}}}
\newcommand{\bfPiddot}{\boldsymbol{\ddot{\Pi}}}
\newcommand{\bfSigmaddot}{\boldsymbol{\ddot{\Sigma}}}
\newcommand{\bfUpsilonddot}{\boldsymbol{\ddot{\Upsilon}}}
\newcommand{\bfPhiddot}{\boldsymbol{\ddot{\Phi}}}
\newcommand{\bfPsiddot}{\boldsymbol{\ddot{\Psi}}}
\newcommand{\bfOmegaddot}{\boldsymbol{\ddot{\Omega}}}

% Basic Latin Lower Hat
\newcommand{\ahat}{\hat{a}}
\newcommand{\bhat}{\hat{b}}
\newcommand{\chat}{\hat{c}}
\newcommand{\dhat}{\hat{d}}
\newcommand{\ehat}{\hat{e}}
\newcommand{\fhat}{\hat{f}}
\newcommand{\ghat}{\hat{g}}
\newcommand{\hhat}{\hat{h}}
\newcommand{\ihat}{\hat{i}}
\newcommand{\jhat}{\hat{j}}
\newcommand{\khat}{\hat{k}}
\newcommand{\lhat}{\hat{l}}
\newcommand{\mhat}{\hat{m}}
\newcommand{\nhat}{\hat{n}}
\newcommand{\ohat}{\hat{o}}
\newcommand{\phat}{\hat{p}}
\newcommand{\qhat}{\hat{q}}
\newcommand{\rhat}{\hat{r}}
\newcommand{\shat}{\hat{s}}
\newcommand{\that}{\hat{t}}
\newcommand{\uhat}{\hat{u}}
\newcommand{\vhat}{\hat{v}}
\newcommand{\what}{\hat{w}}
\newcommand{\xhat}{\hat{x}}
\newcommand{\yhat}{\hat{y}}
\newcommand{\zhat}{\hat{z}}

% Basic Latin Upper Hat
\newcommand{\Ahat}{\hat{A}}
\newcommand{\Bhat}{\hat{B}}
\newcommand{\Chat}{\hat{C}}
\newcommand{\Dhat}{\hat{D}}
\newcommand{\Ehat}{\hat{E}}
\newcommand{\Fhat}{\hat{F}}
\newcommand{\Ghat}{\hat{G}}
\newcommand{\Hhat}{\hat{H}}
\newcommand{\Ihat}{\hat{I}}
\newcommand{\Jhat}{\hat{J}}
\newcommand{\Khat}{\hat{K}}
\newcommand{\Lhat}{\hat{L}}
\newcommand{\Mhat}{\hat{M}}
\newcommand{\Nhat}{\hat{N}}
\newcommand{\Ohat}{\hat{O}}
\newcommand{\Phat}{\hat{P}}
\newcommand{\Qhat}{\hat{Q}}
\newcommand{\Rhat}{\hat{R}}
\newcommand{\Shat}{\hat{S}}
\newcommand{\That}{\hat{T}}
\newcommand{\Uhat}{\hat{U}}
\newcommand{\Vhat}{\hat{V}}
\newcommand{\What}{\hat{W}}
\newcommand{\Xhat}{\hat{X}}
\newcommand{\Yhat}{\hat{Y}}
\newcommand{\Zhat}{\hat{Z}}

% Basic Latin Lower Tilde
\newcommand{\atilde}{\tilde{a}}
\newcommand{\btilde}{\tilde{b}}
\newcommand{\ctilde}{\tilde{c}}
\newcommand{\dtilde}{\tilde{d}}
\newcommand{\etilde}{\tilde{e}}
\newcommand{\ftilde}{\tilde{f}}
\newcommand{\gtilde}{\tilde{g}}
\newcommand{\htilde}{\tilde{h}}
\newcommand{\itilde}{\tilde{i}}
\newcommand{\jtilde}{\tilde{j}}
\newcommand{\ktilde}{\tilde{k}}
\newcommand{\ltilde}{\tilde{l}}
\newcommand{\mtilde}{\tilde{m}}
\newcommand{\ntilde}{\tilde{n}}
\newcommand{\otilde}{\tilde{o}}
\newcommand{\ptilde}{\tilde{p}}
\newcommand{\qtilde}{\tilde{q}}
\newcommand{\rtilde}{\tilde{r}}
\newcommand{\stilde}{\tilde{s}}
\newcommand{\ttilde}{\tilde{t}}
\newcommand{\utilde}{\tilde{u}}
\newcommand{\vtilde}{\tilde{v}}
\newcommand{\wtilde}{\tilde{w}}
\newcommand{\xtilde}{\tilde{x}}
\newcommand{\ytilde}{\tilde{y}}
\newcommand{\ztilde}{\tilde{z}}

% Basic Latin Upper Tilde
\newcommand{\Atilde}{\tilde{A}}
\newcommand{\Btilde}{\tilde{B}}
\newcommand{\Ctilde}{\tilde{C}}
\newcommand{\Dtilde}{\tilde{D}}
\newcommand{\Etilde}{\tilde{E}}
\newcommand{\Ftilde}{\tilde{F}}
\newcommand{\Gtilde}{\tilde{G}}
\newcommand{\Htilde}{\tilde{H}}
\newcommand{\Itilde}{\tilde{I}}
\newcommand{\Jtilde}{\tilde{J}}
\newcommand{\Ktilde}{\tilde{K}}
\newcommand{\Ltilde}{\tilde{L}}
\newcommand{\Mtilde}{\tilde{M}}
\newcommand{\Ntilde}{\tilde{N}}
\newcommand{\Otilde}{\tilde{O}}
\newcommand{\Ptilde}{\tilde{P}}
\newcommand{\Qtilde}{\tilde{Q}}
\newcommand{\Rtilde}{\tilde{R}}
\newcommand{\Stilde}{\tilde{S}}
\newcommand{\Ttilde}{\tilde{T}}
\newcommand{\Utilde}{\tilde{U}}
\newcommand{\Vtilde}{\tilde{V}}
\newcommand{\Wtilde}{\tilde{W}}
\newcommand{\Xtilde}{\tilde{X}}
\newcommand{\Ytilde}{\tilde{Y}}
\newcommand{\Ztilde}{\tilde{Z}}

% Basic Latin Lower Check
\newcommand{\acheck}{\check{a}}
\newcommand{\bcheck}{\check{b}}
\newcommand{\ccheck}{\check{c}}
\newcommand{\dcheck}{\check{d}}
\newcommand{\echeck}{\check{e}}
\newcommand{\fcheck}{\check{f}}
\newcommand{\gcheck}{\check{g}}
\newcommand{\hcheck}{\check{h}}
\newcommand{\icheck}{\check{i}}
\newcommand{\jcheck}{\check{j}}
\newcommand{\kcheck}{\check{k}}
\newcommand{\lcheck}{\check{l}}
\newcommand{\mcheck}{\check{m}}
\newcommand{\ncheck}{\check{n}}
\newcommand{\ocheck}{\check{o}}
\newcommand{\pcheck}{\check{p}}
\newcommand{\qcheck}{\check{q}}
\newcommand{\rcheck}{\check{r}}
\newcommand{\scheck}{\check{s}}
\newcommand{\tcheck}{\check{t}}
\newcommand{\ucheck}{\check{u}}
\newcommand{\vcheck}{\check{v}}
\newcommand{\wcheck}{\check{w}}
\newcommand{\xcheck}{\check{x}}
\newcommand{\ycheck}{\check{y}}
\newcommand{\zcheck}{\check{z}}

% Basic Latin Upper Check
\newcommand{\Acheck}{\check{A}}
\newcommand{\Bcheck}{\check{B}}
\newcommand{\Ccheck}{\check{C}}
\newcommand{\Dcheck}{\check{D}}
\newcommand{\Echeck}{\check{E}}
\newcommand{\Fcheck}{\check{F}}
\newcommand{\Gcheck}{\check{G}}
\newcommand{\Hcheck}{\check{H}}
\newcommand{\Icheck}{\check{I}}
\newcommand{\Jcheck}{\check{J}}
\newcommand{\Kcheck}{\check{K}}
\newcommand{\Lcheck}{\check{L}}
\newcommand{\Mcheck}{\check{M}}
\newcommand{\Ncheck}{\check{N}}
\newcommand{\Ocheck}{\check{O}}
\newcommand{\Pcheck}{\check{P}}
\newcommand{\Qcheck}{\check{Q}}
\newcommand{\Rcheck}{\check{R}}
\newcommand{\Scheck}{\check{S}}
\newcommand{\Tcheck}{\check{T}}
\newcommand{\Ucheck}{\check{U}}
\newcommand{\Vcheck}{\check{V}}
\newcommand{\Wcheck}{\check{W}}
\newcommand{\Xcheck}{\check{X}}
\newcommand{\Ycheck}{\check{Y}}
\newcommand{\Zcheck}{\check{Z}}

% Basic Latin Lower Dot
\newcommand{\adot}{\dot{a}}
\newcommand{\bdot}{\dot{b}}
\newcommand{\edot}{\dot{e}}
\newcommand{\fdot}{\dot{f}}
\newcommand{\gdot}{\dot{g}}
\newcommand{\hdot}{\dot{h}}
\newcommand{\idot}{\dot{i}}
\newcommand{\jdot}{\dot{j}}
\newcommand{\kdot}{\dot{k}}
\newcommand{\ldot}{\dot{l}}
\newcommand{\mdot}{\dot{m}}
\newcommand{\ndot}{\dot{n}}
\newcommand{\pdot}{\dot{p}}
\newcommand{\qdot}{\dot{q}}
\newcommand{\rdot}{\dot{r}}
\newcommand{\sdot}{\dot{s}}
\newcommand{\tdot}{\dot{t}}
\newcommand{\udot}{\dot{u}}
\newcommand{\vdot}{\dot{v}}
\newcommand{\wdot}{\dot{w}}
\newcommand{\xdot}{\dot{x}}
\newcommand{\ydot}{\dot{y}}
\newcommand{\zdot}{\dot{z}}

% Basic Latin Upper Dot
\newcommand{\Adot}{\dot{A}}
\newcommand{\Bdot}{\dot{B}}
\newcommand{\Cdot}{\dot{C}}
\newcommand{\Edot}{\dot{E}}
\newcommand{\Fdot}{\dot{F}}
\newcommand{\Gdot}{\dot{G}}
\newcommand{\Hdot}{\dot{H}}
\newcommand{\Idot}{\dot{I}}
\newcommand{\Jdot}{\dot{J}}
\newcommand{\Kdot}{\dot{K}}
\newcommand{\Ldot}{\dot{L}}
\newcommand{\Mdot}{\dot{M}}
\newcommand{\Ndot}{\dot{N}}
\newcommand{\Odot}{\dot{O}}
\newcommand{\Pdot}{\dot{P}}
\newcommand{\Qdot}{\dot{Q}}
\newcommand{\Rdot}{\dot{R}}
\newcommand{\Sdot}{\dot{S}}
\newcommand{\Tdot}{\dot{T}}
\newcommand{\Udot}{\dot{U}}
\newcommand{\Vdot}{\dot{V}}
\newcommand{\Wdot}{\dot{W}}
\newcommand{\Xdot}{\dot{X}}
\newcommand{\Ydot}{\dot{Y}}
\newcommand{\Zdot}{\dot{Z}}

% Basic Latin Lower DDot
\newcommand{\addot}{\ddot{a}}
\newcommand{\bddot}{\ddot{b}}
\newcommand{\cddot}{\ddot{c}}
\newcommand{\eddot}{\ddot{e}}
\newcommand{\fddot}{\ddot{f}}
\newcommand{\gddot}{\ddot{g}}
\newcommand{\hddot}{\ddot{h}}
\newcommand{\iddot}{\ddot{i}}
\newcommand{\jddot}{\ddot{j}}
\newcommand{\kddot}{\ddot{k}}
\newcommand{\lddot}{\ddot{l}}
\newcommand{\mddot}{\ddot{m}}
\newcommand{\nddot}{\ddot{n}}
\newcommand{\oddot}{\ddot{o}}
\newcommand{\pddot}{\ddot{p}}
\newcommand{\qddot}{\ddot{q}}
\newcommand{\rddot}{\ddot{r}}
\newcommand{\sddot}{\ddot{s}}
\newcommand{\tddot}{\ddot{t}}
\newcommand{\uddot}{\ddot{u}}
\newcommand{\vddot}{\ddot{v}}
\newcommand{\wddot}{\ddot{w}}
\newcommand{\xddot}{\ddot{x}}
\newcommand{\yddot}{\ddot{y}}
\newcommand{\zddot}{\ddot{z}}

% Basic Latin Upper DDot
\newcommand{\Addot}{\ddot{A}}
\newcommand{\Bddot}{\ddot{B}}
\newcommand{\Cddot}{\ddot{C}}
\newcommand{\Dddot}{\ddot{D}}
\newcommand{\Eddot}{\ddot{E}}
\newcommand{\Fddot}{\ddot{F}}
\newcommand{\Gddot}{\ddot{G}}
\newcommand{\Hddot}{\ddot{H}}
\newcommand{\Iddot}{\ddot{I}}
\newcommand{\Jddot}{\ddot{J}}
\newcommand{\Kddot}{\ddot{K}}
\newcommand{\Lddot}{\ddot{L}}
\newcommand{\Mddot}{\ddot{M}}
\newcommand{\Nddot}{\ddot{N}}
\newcommand{\Oddot}{\ddot{O}}
\newcommand{\Pddot}{\ddot{P}}
\newcommand{\Qddot}{\ddot{Q}}
\newcommand{\Rddot}{\ddot{R}}
\newcommand{\Sddot}{\ddot{S}}
\newcommand{\Tddot}{\ddot{T}}
\newcommand{\Uddot}{\ddot{U}}
\newcommand{\Vddot}{\ddot{V}}
\newcommand{\Wddot}{\ddot{W}}
\newcommand{\Xddot}{\ddot{X}}
\newcommand{\Yddot}{\ddot{Y}}
\newcommand{\Zddot}{\ddot{Z}}

% Bold Latin Lower Hat
\newcommand{\bfahat}{\mathbf{\hat{a}}}
\newcommand{\bfbhat}{\mathbf{\hat{b}}}
\newcommand{\bfchat}{\mathbf{\hat{c}}}
\newcommand{\bfdhat}{\mathbf{\hat{d}}}
\newcommand{\bfehat}{\mathbf{\hat{e}}}
\newcommand{\bffhat}{\mathbf{\hat{f}}}
\newcommand{\bfghat}{\mathbf{\hat{g}}}
\newcommand{\bfhhat}{\mathbf{\hat{h}}}
\newcommand{\bfihat}{\mathbf{\hat{i}}}
\newcommand{\bfjhat}{\mathbf{\hat{j}}}
\newcommand{\bfkhat}{\mathbf{\hat{k}}}
\newcommand{\bflhat}{\mathbf{\hat{l}}}
\newcommand{\bfmhat}{\mathbf{\hat{m}}}
\newcommand{\bfnhat}{\mathbf{\hat{n}}}
\newcommand{\bfohat}{\mathbf{\hat{o}}}
\newcommand{\bfphat}{\mathbf{\hat{p}}}
\newcommand{\bfqhat}{\mathbf{\hat{q}}}
\newcommand{\bfrhat}{\mathbf{\hat{r}}}
\newcommand{\bfshat}{\mathbf{\hat{s}}}
\newcommand{\bfthat}{\mathbf{\hat{t}}}
\newcommand{\bfuhat}{\mathbf{\hat{u}}}
\newcommand{\bfvhat}{\mathbf{\hat{v}}}
\newcommand{\bfwhat}{\mathbf{\hat{w}}}
\newcommand{\bfxhat}{\mathbf{\hat{x}}}
\newcommand{\bfyhat}{\mathbf{\hat{y}}}
\newcommand{\bfzhat}{\mathbf{\hat{z}}}

% Bold Latin Upper Hat
\newcommand{\bfAhat}{\mathbf{\hat{A}}}
\newcommand{\bfBhat}{\mathbf{\hat{B}}}
\newcommand{\bfChat}{\mathbf{\hat{C}}}
\newcommand{\bfDhat}{\mathbf{\hat{D}}}
\newcommand{\bfEhat}{\mathbf{\hat{E}}}
\newcommand{\bfFhat}{\mathbf{\hat{F}}}
\newcommand{\bfGhat}{\mathbf{\hat{G}}}
\newcommand{\bfHhat}{\mathbf{\hat{H}}}
\newcommand{\bfIhat}{\mathbf{\hat{I}}}
\newcommand{\bfJhat}{\mathbf{\hat{J}}}
\newcommand{\bfKhat}{\mathbf{\hat{K}}}
\newcommand{\bfLhat}{\mathbf{\hat{L}}}
\newcommand{\bfMhat}{\mathbf{\hat{M}}}
\newcommand{\bfNhat}{\mathbf{\hat{N}}}
\newcommand{\bfOhat}{\mathbf{\hat{O}}}
\newcommand{\bfPhat}{\mathbf{\hat{P}}}
\newcommand{\bfQhat}{\mathbf{\hat{Q}}}
\newcommand{\bfRhat}{\mathbf{\hat{R}}}
\newcommand{\bfShat}{\mathbf{\hat{S}}}
\newcommand{\bfThat}{\mathbf{\hat{T}}}
\newcommand{\bfUhat}{\mathbf{\hat{U}}}
\newcommand{\bfVhat}{\mathbf{\hat{V}}}
\newcommand{\bfWhat}{\mathbf{\hat{W}}}
\newcommand{\bfXhat}{\mathbf{\hat{X}}}
\newcommand{\bfYhat}{\mathbf{\hat{Y}}}
\newcommand{\bfZhat}{\mathbf{\hat{Z}}}

% Bold Latin Lower Tilde
\newcommand{\bfatilde}{\mathbf{\tilde{a}}}
\newcommand{\bfbtilde}{\mathbf{\tilde{b}}}
\newcommand{\bfctilde}{\mathbf{\tilde{c}}}
\newcommand{\bfdtilde}{\mathbf{\tilde{d}}}
\newcommand{\bfetilde}{\mathbf{\tilde{e}}}
\newcommand{\bfftilde}{\mathbf{\tilde{f}}}
\newcommand{\bfgtilde}{\mathbf{\tilde{g}}}
\newcommand{\bfhtilde}{\mathbf{\tilde{h}}}
\newcommand{\bfitilde}{\mathbf{\tilde{i}}}
\newcommand{\bfjtilde}{\mathbf{\tilde{j}}}
\newcommand{\bfktilde}{\mathbf{\tilde{k}}}
\newcommand{\bfltilde}{\mathbf{\tilde{l}}}
\newcommand{\bfmtilde}{\mathbf{\tilde{m}}}
\newcommand{\bfntilde}{\mathbf{\tilde{n}}}
\newcommand{\bfotilde}{\mathbf{\tilde{o}}}
\newcommand{\bfptilde}{\mathbf{\tilde{p}}}
\newcommand{\bfqtilde}{\mathbf{\tilde{q}}}
\newcommand{\bfrtilde}{\mathbf{\tilde{r}}}
\newcommand{\bfstilde}{\mathbf{\tilde{s}}}
\newcommand{\bfttilde}{\mathbf{\tilde{t}}}
\newcommand{\bfutilde}{\mathbf{\tilde{u}}}
\newcommand{\bfvtilde}{\mathbf{\tilde{v}}}
\newcommand{\bfwtilde}{\mathbf{\tilde{w}}}
\newcommand{\bfxtilde}{\mathbf{\tilde{x}}}
\newcommand{\bfytilde}{\mathbf{\tilde{y}}}
\newcommand{\bfztilde}{\mathbf{\tilde{z}}}

% Bold Latin Upper Tilde
\newcommand{\bfAtilde}{\mathbf{\tilde{A}}}
\newcommand{\bfBtilde}{\mathbf{\tilde{B}}}
\newcommand{\bfCtilde}{\mathbf{\tilde{C}}}
\newcommand{\bfDtilde}{\mathbf{\tilde{D}}}
\newcommand{\bfEtilde}{\mathbf{\tilde{E}}}
\newcommand{\bfFtilde}{\mathbf{\tilde{F}}}
\newcommand{\bfGtilde}{\mathbf{\tilde{G}}}
\newcommand{\bfHtilde}{\mathbf{\tilde{H}}}
\newcommand{\bfItilde}{\mathbf{\tilde{I}}}
\newcommand{\bfJtilde}{\mathbf{\tilde{J}}}
\newcommand{\bfKtilde}{\mathbf{\tilde{K}}}
\newcommand{\bfLtilde}{\mathbf{\tilde{L}}}
\newcommand{\bfMtilde}{\mathbf{\tilde{M}}}
\newcommand{\bfNtilde}{\mathbf{\tilde{N}}}
\newcommand{\bfOtilde}{\mathbf{\tilde{O}}}
\newcommand{\bfPtilde}{\mathbf{\tilde{P}}}
\newcommand{\bfQtilde}{\mathbf{\tilde{Q}}}
\newcommand{\bfRtilde}{\mathbf{\tilde{R}}}
\newcommand{\bfStilde}{\mathbf{\tilde{S}}}
\newcommand{\bfTtilde}{\mathbf{\tilde{T}}}
\newcommand{\bfUtilde}{\mathbf{\tilde{U}}}
\newcommand{\bfVtilde}{\mathbf{\tilde{V}}}
\newcommand{\bfWtilde}{\mathbf{\tilde{W}}}
\newcommand{\bfXtilde}{\mathbf{\tilde{X}}}
\newcommand{\bfYtilde}{\mathbf{\tilde{Y}}}
\newcommand{\bfZtilde}{\mathbf{\tilde{Z}}}

% Bold Latin Lower Check
\newcommand{\bfacheck}{\mathbf{\check{a}}}
\newcommand{\bfbcheck}{\mathbf{\check{b}}}
\newcommand{\bfccheck}{\mathbf{\check{c}}}
\newcommand{\bfdcheck}{\mathbf{\check{d}}}
\newcommand{\bfecheck}{\mathbf{\check{e}}}
\newcommand{\bffcheck}{\mathbf{\check{f}}}
\newcommand{\bfgcheck}{\mathbf{\check{g}}}
\newcommand{\bfhcheck}{\mathbf{\check{h}}}
\newcommand{\bficheck}{\mathbf{\check{i}}}
\newcommand{\bfjcheck}{\mathbf{\check{j}}}
\newcommand{\bfkcheck}{\mathbf{\check{k}}}
\newcommand{\bflcheck}{\mathbf{\check{l}}}
\newcommand{\bfmcheck}{\mathbf{\check{m}}}
\newcommand{\bfncheck}{\mathbf{\check{n}}}
\newcommand{\bfocheck}{\mathbf{\check{o}}}
\newcommand{\bfpcheck}{\mathbf{\check{p}}}
\newcommand{\bfqcheck}{\mathbf{\check{q}}}
\newcommand{\bfrcheck}{\mathbf{\check{r}}}
\newcommand{\bfscheck}{\mathbf{\check{s}}}
\newcommand{\bftcheck}{\mathbf{\check{t}}}
\newcommand{\bfucheck}{\mathbf{\check{u}}}
\newcommand{\bfvcheck}{\mathbf{\check{v}}}
\newcommand{\bfwcheck}{\mathbf{\check{w}}}
\newcommand{\bfxcheck}{\mathbf{\check{x}}}
\newcommand{\bfycheck}{\mathbf{\check{y}}}
\newcommand{\bfzcheck}{\mathbf{\check{z}}}

% Bold Latin Upper Check
\newcommand{\bfAcheck}{\mathbf{\check{A}}}
\newcommand{\bfBcheck}{\mathbf{\check{B}}}
\newcommand{\bfCcheck}{\mathbf{\check{C}}}
\newcommand{\bfDcheck}{\mathbf{\check{D}}}
\newcommand{\bfEcheck}{\mathbf{\check{E}}}
\newcommand{\bfFcheck}{\mathbf{\check{F}}}
\newcommand{\bfGcheck}{\mathbf{\check{G}}}
\newcommand{\bfHcheck}{\mathbf{\check{H}}}
\newcommand{\bfIcheck}{\mathbf{\check{I}}}
\newcommand{\bfJcheck}{\mathbf{\check{J}}}
\newcommand{\bfKcheck}{\mathbf{\check{K}}}
\newcommand{\bfLcheck}{\mathbf{\check{L}}}
\newcommand{\bfMcheck}{\mathbf{\check{M}}}
\newcommand{\bfNcheck}{\mathbf{\check{N}}}
\newcommand{\bfOcheck}{\mathbf{\check{O}}}
\newcommand{\bfPcheck}{\mathbf{\check{P}}}
\newcommand{\bfQcheck}{\mathbf{\check{Q}}}
\newcommand{\bfRcheck}{\mathbf{\check{R}}}
\newcommand{\bfScheck}{\mathbf{\check{S}}}
\newcommand{\bfTcheck}{\mathbf{\check{T}}}
\newcommand{\bfUcheck}{\mathbf{\check{U}}}
\newcommand{\bfVcheck}{\mathbf{\check{V}}}
\newcommand{\bfWcheck}{\mathbf{\check{W}}}
\newcommand{\bfXcheck}{\mathbf{\check{X}}}
\newcommand{\bfYcheck}{\mathbf{\check{Y}}}
\newcommand{\bfZcheck}{\mathbf{\check{Z}}}

% Bold Latin Lower Dot
\newcommand{\bfadot}{\mathbf{\dot{a}}}
\newcommand{\bfbdot}{\mathbf{\dot{b}}}
\newcommand{\bfcdot}{\mathbf{\dot{c}}}
\newcommand{\bfddot}{\mathbf{\dot{d}}}
\newcommand{\bfedot}{\mathbf{\dot{e}}}
\newcommand{\bffdot}{\mathbf{\dot{f}}}
\newcommand{\bfgdot}{\mathbf{\dot{g}}}
\newcommand{\bfhdot}{\mathbf{\dot{h}}}
\newcommand{\bfidot}{\mathbf{\dot{i}}}
\newcommand{\bfjdot}{\mathbf{\dot{j}}}
\newcommand{\bfkdot}{\mathbf{\dot{k}}}
\newcommand{\bfldot}{\mathbf{\dot{l}}}
\newcommand{\bfmdot}{\mathbf{\dot{m}}}
\newcommand{\bfndot}{\mathbf{\dot{n}}}
\newcommand{\bfodot}{\mathbf{\dot{o}}}
\newcommand{\bfpdot}{\mathbf{\dot{p}}}
\newcommand{\bfqdot}{\mathbf{\dot{q}}}
\newcommand{\bfrdot}{\mathbf{\dot{r}}}
\newcommand{\bfsdot}{\mathbf{\dot{s}}}
\newcommand{\bftdot}{\mathbf{\dot{t}}}
\newcommand{\bfudot}{\mathbf{\dot{u}}}
\newcommand{\bfvdot}{\mathbf{\dot{v}}}
\newcommand{\bfwdot}{\mathbf{\dot{w}}}
\newcommand{\bfxdot}{\mathbf{\dot{x}}}
\newcommand{\bfydot}{\mathbf{\dot{y}}}
\newcommand{\bfzdot}{\mathbf{\dot{z}}}

% Bold Latin Upper Dot
\newcommand{\bfAdot}{\mathbf{\dot{A}}}
\newcommand{\bfBdot}{\mathbf{\dot{B}}}
\newcommand{\bfCdot}{\mathbf{\dot{C}}}
\newcommand{\bfDdot}{\mathbf{\dot{D}}}
\newcommand{\bfEdot}{\mathbf{\dot{E}}}
\newcommand{\bfFdot}{\mathbf{\dot{F}}}
\newcommand{\bfGdot}{\mathbf{\dot{G}}}
\newcommand{\bfHdot}{\mathbf{\dot{H}}}
\newcommand{\bfIdot}{\mathbf{\dot{I}}}
\newcommand{\bfJdot}{\mathbf{\dot{J}}}
\newcommand{\bfKdot}{\mathbf{\dot{K}}}
\newcommand{\bfLdot}{\mathbf{\dot{L}}}
\newcommand{\bfMdot}{\mathbf{\dot{M}}}
\newcommand{\bfNdot}{\mathbf{\dot{N}}}
\newcommand{\bfOdot}{\mathbf{\dot{O}}}
\newcommand{\bfPdot}{\mathbf{\dot{P}}}
\newcommand{\bfQdot}{\mathbf{\dot{Q}}}
\newcommand{\bfRdot}{\mathbf{\dot{R}}}
\newcommand{\bfSdot}{\mathbf{\dot{S}}}
\newcommand{\bfTdot}{\mathbf{\dot{T}}}
\newcommand{\bfUdot}{\mathbf{\dot{U}}}
\newcommand{\bfVdot}{\mathbf{\dot{V}}}
\newcommand{\bfWdot}{\mathbf{\dot{W}}}
\newcommand{\bfXdot}{\mathbf{\dot{X}}}
\newcommand{\bfYdot}{\mathbf{\dot{Y}}}
\newcommand{\bfZdot}{\mathbf{\dot{Z}}}

% Bold Latin Lower DDot
\newcommand{\bfaddot}{\mathbf{\ddot{a}}}
\newcommand{\bfbddot}{\mathbf{\ddot{b}}}
\newcommand{\bfcddot}{\mathbf{\ddot{c}}}
\newcommand{\bfdddot}{\mathbf{\ddot{d}}}
\newcommand{\bfeddot}{\mathbf{\ddot{e}}}
\newcommand{\bffddot}{\mathbf{\ddot{f}}}
\newcommand{\bfgddot}{\mathbf{\ddot{g}}}
\newcommand{\bfhddot}{\mathbf{\ddot{h}}}
\newcommand{\bfiddot}{\mathbf{\ddot{i}}}
\newcommand{\bfjddot}{\mathbf{\ddot{j}}}
\newcommand{\bfkddot}{\mathbf{\ddot{k}}}
\newcommand{\bflddot}{\mathbf{\ddot{l}}}
\newcommand{\bfmddot}{\mathbf{\ddot{m}}}
\newcommand{\bfnddot}{\mathbf{\ddot{n}}}
\newcommand{\bfoddot}{\mathbf{\ddot{o}}}
\newcommand{\bfpddot}{\mathbf{\ddot{p}}}
\newcommand{\bfqddot}{\mathbf{\ddot{q}}}
\newcommand{\bfrddot}{\mathbf{\ddot{r}}}
\newcommand{\bfsddot}{\mathbf{\ddot{s}}}
\newcommand{\bftddot}{\mathbf{\ddot{t}}}
\newcommand{\bfuddot}{\mathbf{\ddot{u}}}
\newcommand{\bfvddot}{\mathbf{\ddot{v}}}
\newcommand{\bfwddot}{\mathbf{\ddot{w}}}
\newcommand{\bfxddot}{\mathbf{\ddot{x}}}
\newcommand{\bfyddot}{\mathbf{\ddot{y}}}
\newcommand{\bfzddot}{\mathbf{\ddot{z}}}

% Bold Latin Upper DDot
\newcommand{\bfAddot}{\mathbf{\ddot{A}}}
\newcommand{\bfBddot}{\mathbf{\ddot{B}}}
\newcommand{\bfCddot}{\mathbf{\ddot{C}}}
\newcommand{\bfDddot}{\mathbf{\ddot{D}}}
\newcommand{\bfEddot}{\mathbf{\ddot{E}}}
\newcommand{\bfFddot}{\mathbf{\ddot{F}}}
\newcommand{\bfGddot}{\mathbf{\ddot{G}}}
\newcommand{\bfHddot}{\mathbf{\ddot{H}}}
\newcommand{\bfIddot}{\mathbf{\ddot{I}}}
\newcommand{\bfJddot}{\mathbf{\ddot{J}}}
\newcommand{\bfKddot}{\mathbf{\ddot{K}}}
\newcommand{\bfLddot}{\mathbf{\ddot{L}}}
\newcommand{\bfMddot}{\mathbf{\ddot{M}}}
\newcommand{\bfNddot}{\mathbf{\ddot{N}}}
\newcommand{\bfOddot}{\mathbf{\ddot{O}}}
\newcommand{\bfPddot}{\mathbf{\ddot{P}}}
\newcommand{\bfQddot}{\mathbf{\ddot{Q}}}
\newcommand{\bfRddot}{\mathbf{\ddot{R}}}
\newcommand{\bfSddot}{\mathbf{\ddot{S}}}
\newcommand{\bfTddot}{\mathbf{\ddot{T}}}
\newcommand{\bfUddot}{\mathbf{\ddot{U}}}
\newcommand{\bfVddot}{\mathbf{\ddot{V}}}
\newcommand{\bfWddot}{\mathbf{\ddot{W}}}
\newcommand{\bfXddot}{\mathbf{\ddot{X}}}
\newcommand{\bfYddot}{\mathbf{\ddot{Y}}}
\newcommand{\bfZddot}{\mathbf{\ddot{Z}}}

% Bold Latin Lower
\newcommand{\bfa}{\mathbf{a}}
\newcommand{\bfb}{\mathbf{b}}
\newcommand{\bfc}{\mathbf{c}}
\newcommand{\bfd}{\mathbf{d}}
\newcommand{\bfe}{\mathbf{e}}
\newcommand{\bff}{\mathbf{f}}
\newcommand{\bfg}{\mathbf{g}}
\newcommand{\bfh}{\mathbf{h}}
\newcommand{\bfi}{\mathbf{i}}
\newcommand{\bfj}{\mathbf{j}}
\newcommand{\bfk}{\mathbf{k}}
\newcommand{\bfl}{\mathbf{l}}
\newcommand{\bfm}{\mathbf{m}}
\newcommand{\bfn}{\mathbf{n}}
\newcommand{\bfo}{\mathbf{o}}
\newcommand{\bfp}{\mathbf{p}}
\newcommand{\bfq}{\mathbf{q}}
\newcommand{\bfr}{\mathbf{r}}
\newcommand{\bfs}{\mathbf{s}}
\newcommand{\bft}{\mathbf{t}}
\newcommand{\bfu}{\mathbf{u}}
\newcommand{\bfv}{\mathbf{v}}
\newcommand{\bfw}{\mathbf{w}}
\newcommand{\bfx}{\mathbf{x}}
\newcommand{\bfy}{\mathbf{y}}
\newcommand{\bfz}{\mathbf{z}}

% Bold Latin Upper
\newcommand{\bfA}{\mathbf{A}}
\newcommand{\bfB}{\mathbf{B}}
\newcommand{\bfC}{\mathbf{C}}
\newcommand{\bfD}{\mathbf{D}}
\newcommand{\bfE}{\mathbf{E}}
\newcommand{\bfF}{\mathbf{F}}
\newcommand{\bfG}{\mathbf{G}}
\newcommand{\bfH}{\mathbf{H}}
\newcommand{\bfI}{\mathbf{I}}
\newcommand{\bfJ}{\mathbf{J}}
\newcommand{\bfK}{\mathbf{K}}
\newcommand{\bfL}{\mathbf{L}}
\newcommand{\bfM}{\mathbf{M}}
\newcommand{\bfN}{\mathbf{N}}
\newcommand{\bfO}{\mathbf{O}}
\newcommand{\bfP}{\mathbf{P}}
\newcommand{\bfQ}{\mathbf{Q}}
\newcommand{\bfR}{\mathbf{R}}
\newcommand{\bfS}{\mathbf{S}}
\newcommand{\bfT}{\mathbf{T}}
\newcommand{\bfU}{\mathbf{U}}
\newcommand{\bfV}{\mathbf{V}}
\newcommand{\bfW}{\mathbf{W}}
\newcommand{\bfX}{\mathbf{X}}
\newcommand{\bfY}{\mathbf{Y}}
\newcommand{\bfZ}{\mathbf{Z}}

% Others
\newcommand{\Rc}{\textsuperscript{\textregistered}}
\newcommand{\nline}{\mbox{ }\newline}
\newcommand{\diag}{\textrm{diag}}
\newcommand{\dlim}{\stackrel{D}{\rightarrow}}
\newcommand{\plim}{\stackrel{P}{\rightarrow}}
\newcommand{\eye}{\mathbf{I}}
\newcommand{\tr}{\textrm{tr}}
\newcommand{\var}{\textrm{var}}
\newcommand{\varhat}{\widehat{\textrm{var}}}
\newcommand{\vech}{\textrm{vech}}
\newcommand{\grad}{\nabla_{\bftheta}}
\newcommand{\hess}{\nabla_{\bftheta}^2}
\newcommand{\PD}[2]{\frac{\partial #1}{\partial #2}}
\newcommand{\PDD}[2]{\frac{\partial^2 #1}{\partial #2^2}}
\newcommand{\PDC}[3]{\frac{\partial^2 #1}{\partial #2 \partial #3}}
\newcommand{\tablehere}[1]{
\begin{center}
\underline{Insert Table #1 About Here}\\
\end{center}
}
\newcommand{\figurehere}[1]{
\begin{center}
\underline{Insert Figure #1 About Here}\\
\end{center}
}
\newcommand{\tableshere}[1]{
\begin{center}
\underline{Insert Tables #1 About Here}\\
\end{center}
}
\newcommand{\figureshere}[1]{
\begin{center}
\underline{Insert Figures #1 About Here}\\
\end{center}
}

\lhead{FIT PROPENSITY} \rhead{\thepage} \lfoot{} \cfoot{} \rfoot{}
\renewcommand{\headrulewidth}{0pt} \setlength{\headheight}{15pt}
\setlength\parindent{24pt}

\fancypagestyle{firststyle}{ \lhead{Running Head: FIT PROPENSITY} \rhead{}}

\thispagestyle{firststyle}

\mbox{ }

\vspace{1.5in}

\begin{center}
\textsc{Parsimony in Model Selection: Tools for Assessing Fit Propensity}
\end{center}

\vspace{1.25in}

\begin{center}\textsc{
Carl F. Falk\textsuperscript{*}\\
\nline
McGill University\\
\nline
\nline
Michael Muthukrishna\textsuperscript{*}\\
\nline
London School of Economics and Political Science (LSE)\\
}
\end{center}

\vfill

\noindent 

\textsuperscript{*}The authors contributed equally to this manuscript.
Author order is alphabetical.

\nline

The authors would like to thank Victoria Savalei for her inspiration and
early input on this project. We acknowledge the support of the Natural
Science and Engineering Research Council of Canada (NSERC), (funding
reference number RGPIN-2018-05357 and DGECR-2018-00083). Cette recherche
a ete financee par le Conseil de recherches en sciences naturelles et en
genie du Canada (CRSNG), {[}numero de reference RGPIN-2018-05357{]}. A preprint of an earlier version of the manuscript is available at: https://arxiv.org/abs/2007.03699.
Address all correspondence to: Carl F. Falk, Department of Psychology,
McGill University, 2001 McGill College, 7th Floor, Montreal, QC H3A 1G1,
Canada. Email:
\href{mailto:carl.falk@mcgill.ca}{\nolinkurl{carl.falk@mcgill.ca}}.

\newpage

\setcounter{page}{1}

\begin{center}
\textsc{Parsimony in Model Selection: Tools for Assessing Fit Propensity}
\end{center}

\begin{center}
\textbf{Abstract}
\end{center}

Theories can be represented as statistical models for empirical testing.
There is a vast literature on model selection and multimodel inference
that focuses on how to assess which statistical model, and therefore
which theory, best fits the available data. For example, given some
data, one can compare models on various information criterion or other
fit statistics. However, what these indices fail to capture is the full
range of counterfactuals. That is, some models may fit the given data
better not because they represent a more correct theory, but simply
because these models have more \emph{fit propensity} - a tendency to fit
a wider range of data, even nonsensical data, better. Current approaches
fall short in considering the principle of parsimony (Occam's Razor),
often equating it with the number of model parameters. Here we offer a
toolkit for researchers to better study and understand parsimony through
the fit propensity of Structural Equation Models. We provide an \emph{R}
package (\emph{ockhamSEM}) built on the popular \emph{lavaan} package.
To illustrate the importance of evaluating fit propensity, we use
\emph{ockhamSEM} to investigate the factor structure of the Rosenberg
Self-Esteem Scale.

\textbf{Keywords:} fit indices, parsimony, model fit, structural
equation modeling, formal theory, SEM

\newpage 

\doublespace

\section{Introduction}

Theories, no matter how beautiful, live and die on the back of data.
Structural equation modeling offers a flexible framework for
statistically representing complex theories (Bollen \& Pearl, 2013;
Grace \& Bollen, 2008). Given a choice between two or more theoretically
plausible structural equation models, the process of model selection and
multimodel inference (Burnham \& Anderson, 2002) typically involves
asking which model is more consistent with the available empirical data.
For example, there are many instances in psychological research where a
broad (multifaceted) construct is defined, a test is created, and
further psychometric work indicates that a multidimensional model fits
better than a model that measures a single dimension. Some examples
include the number and configuration of possible method factors on
scales that include reverse-worded items (Reise, Kim, Mansolf, \&
Widaman, 2016) or whether a random intercept model should be used to
model acquiescence bias (Savalei \& Falk, 2014), the tradeoff between a
correlated factor, hierarchical factor, and bifactor models for
constructs such as self-compassion (Neff, Whittaker, \& Karl, 2017),
alexithymia (Reise, Bonifay, \& Haviland, 2013), health outcomes (Reise,
Morizot, \& Hays, 2007), and so on. In all cases, debates continue over
which model is most correct. What is often overlooked is the
counterfactual - that a model may not fit the empirical data better
because it is a better description of reality, but simply because it has
a tendency to fit \emph{any data} better. That is, what is often
overlooked is parsimony.

Occam's razor, or the principle of parsimony, is familiar to most
scientists. As we teach our students: given the choice between two
equally fitting models, all else being equal it is generally preferable
to choose the simpler, or more parsimonious, model. What is less well
understood is how one might quantify parsimony. One promising approach
is the concept of model \emph{fit propensity} (FP; Preacher, 2006) or
\emph{complexity} (Myung, Pitt, \& Kim, 2005; Pitt, Myung, \& Zhang,
2002). Here we will use \emph{fit propensity} to avoid confusion with
other uses of the term \emph{complex}. Fit propensity is sometimes
described as the ``complement'' of parsimony (Preacher, 2006, p. 230).
The basic idea behind fit propensity is that some models will simply do
a better job of fitting a wider range of data. These models are less
parsimonious. Thus, the process of model selection needs to consider not
just model fit, but fit propensity. The ideal theoretically derived
model will have both better fit and lower fit propensity than a
competing model. But in practice, there is likely to be a tension
between fit and fit propensity. In other words, for a model to be both
useful and generalizable, a balance must be struck between fitting real
data, and parsimony in not also fitting random data (fit propensity can be defined as the propensity to fit random data; Bonifay, 2015;
Cudeck \& Browne, 1983; Marsh \& Balla, 1994; Reise et al., 2013).

Parsimony is sometimes described as a function of degrees of freedom.
For example, Marsh \& Balla (1994) defined parsimony as ``the ratio of
degrees of freedom in the model being tested and degrees of freedom in
the null model (James et al., 1982; Mulaik et al., 1989)" (p.~188). It
is thus tempting to equate parsimony with the degrees of freedom of a
model such that fewer estimated parameters (and higher \emph{df})
corresponds to more parsimony. However, it is possible to have models
with the same number of estimated parameters, but where one has better
propensity to fit random data (Bonifay \& Cai, 2017; Preacher, 2006).
Indeed, a model may even have more estimated parameters than an
alternative, but have lower fit propensity and therefore more parsimony
(Pearl \& Verma, 1995). The configuration of the model (number of latent
factors, paths among variables) and functional form of relationships
among variables also affects fit propensity. In sum, prior research has
identified both the number of estimated parameters and the functional form
or configuration of the model as both contributing to fit propensity
(Myung et al., 2005; Preacher, 2006). Thus, fit indices that
adjust for degrees of freedom, such as Tucker-Lewis Index (TLI; Tucker
\& Lewis, 1973), Comparative Fit Index (CFI; Bentler, 1990), and Root
Mean Square Error of Approximation (RMSEA; Steiger \& Lind, 1980) or
commonly used information criterion, such as AIC and BIC that have
adjustments based on the number of estimated parameters are coarse in
how they treat the role of fit propensity in model selection.

\subsection{Fit Propensity and the Rosenberg Self-Esteem Scale}

To understand the importance of fit propensity, consider the Rosenberg
Self-Esteem Scale (RSES; Rosenberg, 1965). The RSES is perhaps the most
widely used self-report instrument for the measurement of self-esteem.
It contains ten 5-point Likert-type items. The RSES is often used by
applied researchers to represent a single construct: self-esteem. Higher
scores indicate higher self-esteem for five positively keyed items
(items 1, 2, 4, 6, and 7), and five negatively worded or reverse keyed
items (items 3, 5, 8, 9, and 10). The RSES is regularly used, but is the
subject of ongoing investigations to examine the confirmatory factor
analysis models that may represent it; a single factor model rarely fits
RSES data adequately. In a recent example, Donnellan, Ackerman, \&
Brecheen (2016) fit ten different models to RSES data (\(N=1,127\)). Of
these, three models stood out as having superior fit: 1) A global factor
with correlated residuals among positively and negatively worded
items\footnote{With one residual fixed to zero for identification.}; 2)
A bifactor model with method factors for positively and negatively
worded items; and 3) The same bifactor model, but with correlated method
factors (Figure \ref{fig:RSES}). For illustration, we replicated the
original analyses using \emph{lavaan} (Rosseel, 2012), and results for
these models and a single factor model are presented in Table
\ref{tbl:RSESfit}.\footnote{To account for ordered categorical data,
  maximum likelihood with robust corrections (i.e., estimator=``MLR'')
  was employed here and in the original paper.}

\begin{table}
\caption{Rosenberg Self-Esteem Model Fit}
\label{tbl:RSESfit}
\begin{center}
\begin{tabular}{lccccccccc}
\toprule
Model & $\chi^2$ & $df$ & TLI & CFI & RMSEA & AIC & BIC & SIC & RMSR\\ \midrule
1. Correlated Residual & 60.59 & 16 & 0.97 & 0.99 & 0.05 & 26343 & 26589 & 26621 & 0.02\\
2. Bifactor& 154.88 & 25 & 0.94 & 0.97 & 0.07 & 26437 & 26638 & 26652 & 0.03\\
3. Correlated Bifactor& 135.09 & 24 & 0.95 & 0.97 & 0.06 & 26413 & 26619 & 26632 & 0.02\\
4. Single Factor& 872.91 & 35 & 0.72 & 0.78 & 0.15 & 27346 & 27497 & 27508 & 0.08\\
\bottomrule
\end{tabular}\\
\end{center}
\end{table}

\begin{figure}[H]
\begin{center}
\caption{Rosenberg Self-Esteem Models}  \vskip10pt
\label{fig:RSES}

\begin{tikzpicture}[>=stealth]
\node[ov] (v1) at (0,0) {$V_1$};
\foreach \x in {2,...,10} {
  \def\X{\the\numexpr\x-1}
  \node[ov] (v\x) [right=5mm of v\X]  {$V_{\x}$};
}
\node[lv] (F1) at (5.5,2.25) {$\eta_1$};
\foreach \x in {1,...,10} \node (e\x) [below=3mm of v\x][scale=0.75, circle, draw,  fill=gray!20, drop shadow] {};
\path[->, thick] (F1) edge node[scale=0.75, left] {} (v1.north);
\foreach \x in {2,...,10} \path[->, thick] (F1) edge node[scale=0.75, left] {} (v\x.north);
\path [<->] (F1) edge [loop above] node [scale=0.75] {} (F1);
\foreach \x in {1,...,10} \path [<->] (e\x) edge [loop below] node [scale=0.75] {} (e\x);
\foreach \x in {1,...,10} \path [->, thick] (e\x) edge node [scale=0.75] {} (v\x);
\path[<->] (e2.south) edge [bend left=35] node[left,scale=0.8] {} (e1.south);
\path[<->] (e4.south) edge [bend left=35] node[left,scale=0.8] {} (e1.south);
\path[<->] (e6.south) edge [bend left=35] node[left,scale=0.8] {} (e1.south);
\path[<->] (e7.south) edge [bend left=35] node[left,scale=0.8] {} (e1.south);
\path[<->] (e4.south) edge [bend left=35] node[left,scale=0.8] {} (e2.south);
\path[<->] (e6.south) edge [bend left=35] node[left,scale=0.8] {} (e2.south);
\path[<->] (e7.south) edge [bend left=35] node[left,scale=0.8] {} (e2.south);
\path[<->] (e6.south) edge [bend left=35] node[left,scale=0.8] {} (e4.south);
\path[<->] (e7.south) edge [bend left=35] node[left,scale=0.8] {} (e4.south);
\path[<->] (e7.south) edge [bend left=35] node[left,scale=0.8] {} (e6.south);

\path[<->] (e5.south) edge [bend left=35] node[left,scale=0.8] {} (e3.south);
\path[<->] (e8.south) edge [bend left=35] node[left,scale=0.8] {} (e3.south);
\path[<->] (e9.south) edge [bend left=35] node[left,scale=0.8] {} (e3.south);
\path[<->] (e10.south) edge [bend left=35] node[left,scale=0.8] {} (e3.south);
\path[<->] (e9.south) edge [bend left=35] node[left,scale=0.8] {} (e5.south);
\path[<->] (e10.south) edge [bend left=35] node[left,scale=0.8] {} (e5.south);
\path[<->] (e9.south) edge [bend left=35] node[left,scale=0.8] {} (e8.south);
\path[<->] (e10.south) edge [bend left=35] node[left,scale=0.8] {} (e8.south);
\path[<->] (e10.south) edge [bend left=35] node[left,scale=0.8] {} (e9.south);

\node[above,font=\large\bfseries] at (current bounding box.north) {Correlated Residual Model};
\end{tikzpicture}

\begin{tikzpicture}[>=stealth]
\node[ov] (v1) at (0,0) {$V_1$};
\foreach \x in {2,...,10} {
  \def\X{\the\numexpr\x-1}
  \node[ov] (v\x) [right=5mm of v\X]  {$V_{\x}$};
}
\node[lv] (F1) at (2.5,-2.5) {$\eta_2$};
\node[lv] (F2) at (8.65,-2.5) {$\eta_3$};
\node[lv] (F3) at (5.5,2.5) {$\eta_1$};
\foreach \x in {1,...,10} \node (e\x) [below=3mm of v\x][scale=0.75, circle, draw,  fill=gray!20, drop shadow] {};
\path[->, thick] (F1) edge node[scale=0.75, left] {} (v1.south);
\path[->, thick] (F2) edge node[scale=0.75, left] {} (v3.south);
\path[->, thick] (F3) edge node[scale=0.75, left] {} (v1.north);
\foreach \x in {1,2,4,6,7} \path[->, thick] (F1) edge node[scale=0.75, left] {} (v\x.south);
\foreach \x in {5,8,9,10} \path[->, thick] (F2) edge node[scale=0.75, left] {} (v\x.south);
\foreach \x in {2,...,10} \path[->, thick] (F3) edge node[scale=0.75, left] {} (v\x.north);
\path [<->] (F1) edge [loop below] node [scale=0.75] {} (F1);
\path [<->] (F2) edge [loop below] node [scale=0.75] {} (F2);
\path [<->] (F3) edge [loop above] node [scale=0.75] {} (F3);
\foreach \x in {1,...,10} \path [<->] (e\x) edge [loop below] node [scale=0.75] {} (e\x);
\foreach \x in {1,...,10} \path [->, thick] (e\x) edge node [scale=0.75] {} (v\x);
\node[above,font=\large\bfseries] at (current bounding box.north) {Bifactor Model};
\end{tikzpicture}

\begin{tikzpicture}[>=stealth]
\node[ov] (v1) at (0,0) {$V_1$};
\foreach \x in {2,...,10} {
  \def\X{\the\numexpr\x-1}
  \node[ov] (v\x) [right=5mm of v\X]  {$V_{\x}$};
}
\node[lv] (F1) at (2.5,-2.5) {$\eta_2$};
\node[lv] (F2) at (8.65,-2.5) {$\eta_3$};
\node[lv] (F3) at (5.5,2.5) {$\eta_1$};
\foreach \x in {1,...,10} \node (e\x) [below=3mm of v\x][scale=0.75, circle, draw,  fill=gray!20, drop shadow] {};
\path[->, thick] (F1) edge node[scale=0.75, left] {} (v1.south);
\path[->, thick] (F2) edge node[scale=0.75, left] {} (v3.south);
\path[->, thick] (F3) edge node[scale=0.75, left] {} (v1.north);
\foreach \x in {1,2,4,6,7} \path[->, thick] (F1) edge node[scale=0.75, left] {} (v\x.south);
\foreach \x in {5,8,9,10} \path[->, thick] (F2) edge node[scale=0.75, left] {} (v\x.south);
\foreach \x in {2,...,10} \path[->, thick] (F3) edge node[scale=0.75, left] {} (v\x.north);
\path [<->] (F1) edge [loop below] node [scale=0.75] {} (F1);
\path [<->] (F2) edge [loop below] node [scale=0.75] {} (F2);
\path [<->] (F3) edge [loop above] node [scale=0.75] {} (F3);
\path[<->] (F2.south) edge [bend left=35] node[left,scale=0.8] {} (F1.south);
\foreach \x in {1,...,10} \path [<->] (e\x) edge [loop below] node [scale=0.75] {} (e\x);
\foreach \x in {1,...,10} \path [->, thick] (e\x) edge node [scale=0.75] {} (v\x);
\node[above,font=\large\bfseries] at (current bounding box.north) {Bifactor Model With Correlated Method Factors};
\end{tikzpicture}

\end{center}
\end{figure}

On the one hand, such well-fitting models may make intuitive sense. All
three models account for additional dependencies beyond a single factor,
and may be appropriate to the extent that positively worded items share
some dependency, as do negatively worded items. On the other hand, one
might question whether these models fit for other reasons. Are they
parsimonious? The correlated residual model essentially accomplishes a
similar task as the bifactor model in modeling dependencies among
similar items, but does so with even more additional model parameters. Does this come with a cost to fit propensity?
Indeed, from a traditional standpoint, these models have the most
estimated parameters of all ten models examined: 39, 30, and 31,
respectively, with only 20 for the single factor model. The original
article primarily considered aforementioned fit indices that make
adjustments based on degrees of freedom or the number of estimated
parameters: TLI, CFI, RMSEA, AIC, and BIC. The only other information
regarding model fit are the chi-square test of fit, and root mean square
residuals (RMSR) - a transformation of the difference between sample
covariances and recovery of covariances by the model. In all cases,
there is either no adjustment for parsimony or only a coarse-grained
adjustment for degrees of freedom.

Bonifay, Lane, \& Reise (2017) argued that the bifactor model may be
good at fitting random noise--that it lacks parsimony. For instance,
Bonifay and Cai (2017) found that a bifactor model with two uncorrelated
method factors had higher fit propensity in general than two hierarchical models
with discrete latent variables with the same number of parameters. In also examining the RSES, Reise, Kim,
Manslof, \& Widaman (2016) found that the bifactor model with
uncorrelated method factors helped explain inconsistent response
patterns, but that a single factor model was sufficient for the majority
of participants. Note that the fit propensity of the additional models
considered by Donnellan et al. (2016) have not been studied. One might
then also wonder--how much more fit propensity does a correlated
residual model have above and beyond a bifactor model? Or, does adding a
single correlation among method factors substantially change fit
propensity? How does the fit propensity of such models compare to a
single factor model? Does fit propensity correlate with number of parameters? Is it possible that such models tend to fit the
data well, not because they are close approximations of reality, but
that such models tend to fit any data, even random data, very well? And
even more broadly, does relative fit propensity depend on which fit
index is examined?

\subsection{Testing Fit Propensity}

The types of questions we ask above provide clues about parsimony that
are not easily answered by number of parameters or traditional fit indices. They can, however,
be understood through a study of fit propensity.\footnote{In our
  replication, an additional fit index, Stochastic Information
  Complexity (SIC; Hansen \& Yu, 2001), is reported and that could be
  used for adjustment of model fit that is more in line with fit
  propensity (Bonifay \& Cai, 2017; Preacher, 2006), but as we discuss
  later in this manuscript, does not immediately provide intuitive
  information regarding fit propensity.} A popular method of studying
fit propensity requires repeated generation of random data from a data
space and fitting the models of interest (Preacher, 2006). Information
regarding model fit can then be recorded over a large number of
replications and summarized to provide a sense of how well the models
fit such random data. Preacher (2006) introduced the concept of fit
propensity to SEM over a decade ago. Further research has been limited,
perhaps in part due to a lack of easy to use and efficient software
tools for evaluating fit propensity. Preacher's (2003) original code was
written in FORTRAN and had a few limitations, such as a Markov Chain
Monte Carlo (MCMC) algorithm that took a long time to generate random
correlation matrices, restriction to only positive correlations, use of
a lesser-known software program for fitting models (RAMONA 4.0 for DOS;
Browne \& Mels, 1990 as cited in Preacher, 2006), and support for few
fit indices (only RMSR was studied)\footnote{We thank Kris Preacher for
  graciously providing us this FORTRAN code, which also appears in his
  dissertation.}.

We aim to support further researchers in considering fit propensity of
their models by providing an \emph{R} package: \emph{ockhamSEM}. The
\emph{ockhamSEM} package offers easy-to-use and highly flexible software
built on the popular \emph{lavaan} (Rosseel, 2012) package. We hope that
\emph{ockhamSEM} will be used for the study of fit propensity by applied
researchers investigating models of interest, for classroom
demonstrations, or the further study of fit propensity itself and
related methodological challenges by quantitative methodologists.

Investigating fit propensity requires generating random correlation
matrices, which are computationally intensive. The \emph{ockhamSEM}
package provides several innovations in terms of both computational
efficiency and the reporting of fit propensity. In particular, random
correlation matrices can be generated using the \emph{onion} method by
Lewandowski and colleagues (2009), as well as Preacher's original MCMC
algorithm. The onion method exploits known properties of elliptically contoured distributions applied to a k-dimensional hypersphere to provide a space of correlation matrices that can be sampled. This sampling results in the generation of correlation matrices much faster than the MCMC method, which involves iteratively generated random draws, where as with many MCMC methods, most are discarded. Using the onion method, thousands of large correlation matrices can be generated in seconds. We discuss these methods in further detail with additional resources in the Appendix.

Calculations can be performed in parallel using the multiple
processing cores common in modern personal computers and computing
clusters. Random correlation matrices can be restricted to all positive
correlations, or both positive and negative correlations (indeed, other
arbitrary restrictions can also be implemented). We also provide
support for the full range of fit indices available from \emph{lavaan}.
Finally, additional numerical and graphical summaries are provided,
going beyond those originally presented by Preacher (2006).

Our work is related to some recent research on fit propensity and model
similarity. In particular, Bonifay \& Cai (2017) describe methods for
studying the fit propensity of item response models with categorical
observed variables. Given the unification of item response models and
SEM under a unified latent variable modeling framework (e.g., Skrondal
\& Rabe-Hesketh, 2004), this work is related to the present research.
However, it does not address continuous observed variables and none of
the underlying code was provided. We take the ``ameoba" plots presented
by Bonifay \& Cai (2017) as inspiration for some vizualizations we
present later in this paper. In addition, Lai and colleagues (2017)
address methods for examining model similarity using mostly scatter
plots and line graphs of fit indices, and may be helpful for visualizing
whether some models are equivalent or nested (but see Bentler \&
Satorra, 2010). These authors mention difficulty in generating data from
random correlation matrices, and instead opt for data generation from a
restricted space that is a mixture of the correlation matrices implied
by two competing models. Thus, model similarity rather than fit
propensity was the main focus of this previous work.

This paper is organized into the following sections. The first section
provides a brief description of our implementation of the R code. We
then illustrate concepts of fit propensity and basic features of the
code in the context of several initial examples from Preacher (2006),
and the RSES example. Finally, we conclude with a discussion of
additional innovations and alternative ways to compare models.

\section{Illustrative Examples}\label{illustrative-examples}

We present three examples to illustrate the basic procedure and concepts
used to study fit propensity, including visualization and summaries of
results (Table \ref{tbl:examples}). The first two examples expand upon
those initially presented by Preacher (2006). We note that while the
general pattern of results remains similar in our implementation, there
may be minor discrepancies for a number of reasons.\footnote{Different
  SEM program with different default estimation options, different
  handling of non-converging models (\emph{lavaan} does not allow
  calculation of some fit indices), etc. Preacher (2006) also used
  ordinary least squares for estimation, whereas we used maximum
  likelihood. For this paper, we used \emph{lavaan} version 0.6-5 and R
  version 3.6.0.} The final example concentrates on the debate around
the RSES and what a study of fit propensity can provide. The first
example is fully illustrated in-text with complete R code. The code for
additional examples is available in the Supplementary Materials.

\begin{table}
\caption{Overview of Examples}
\label{tbl:examples}
\begin{center}
\begin{tabular}{ccp{1.5in}p{3in}}
\toprule
Example & \# Models & Model Description & Main Purpose/Illustrated Features\\ \midrule
A & 2 & Two 3-variable models & Basic use of code, algorithms for correlation matrix generation, parallel processing, equal \textit{df} models, empirical ECDF plots, quantiles\\
B & 2 & Factor and simplex models & Positive vs. negative correlations, equal \textit{df} models, model convergence\\
C & 4 & Four RSES Models & Other fit indices (CFI, RMSEA, TLI), Euler plots, saving of correlation matrices and fitted models\\
\bottomrule
\end{tabular}\\
\end{center}
\end{table}

\subsection{Example A: Fit Propensity
Basics}\label{example-a-fit-propensity-basics}

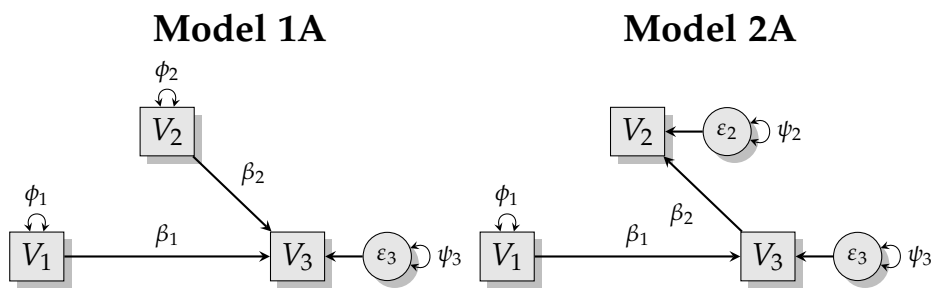
\begin{figure}[h]
\begin{center}
\caption{Two 3-variable models.}  \vskip10pt
\label{fig:figA}
\begin{tikzpicture}[auto, >=stealth,
    error/.style={edge from parent/.style={draw, <-}},
    level 1/.style={level distance=7mm},
    level 2/.style={level distance=4mm}]
    \node (V1) [rectangle, draw,  fill=gray!20, drop shadow] {$V_1$};
    \node (V2) [above right=of V1] [rectangle, draw,  fill=gray!20, drop shadow] {$V_2$};
    \node (V3)[below right=of V2] [rectangle, draw,  fill=gray!20, drop shadow] {$V_3$};
    \node (e3) [right =5mm of V3] [scale=0.75, circle, draw, fill=gray!20, drop shadow] {$\varepsilon_3$};
    \path [->, thick] (V1) edge node [scale=0.75] {$\beta_1$} (V3);
    \path [->, thick] (V2) edge node [scale=0.75] {$\beta_2$} (V3);
    \path [->, thick] (e3) edge node [scale=0.75] {} (V3);
    \path [<->] (V1) edge [loop above] node [scale=0.75] {$\phi_1$} (V1);
    \path [<->] (V2) edge [loop above] node [scale=0.75] {$\phi_2$} (V2);
    \draw [<->] (e3) to [out=20, in=340, looseness=4] node [scale=0.75] {$\psi_3$} (e3);
    \node[above,font=\large\bfseries] at (current bounding box.north) {Model 1A};
\end{tikzpicture}
\begin{tikzpicture}[auto, >=stealth,
    error/.style={edge from parent/.style={draw, <-}},
    level 1/.style={level distance=7mm},
    level 2/.style={level distance=4mm}]
    \node (V1) [rectangle, draw,  fill=gray!20, drop shadow] {$V_1$};
    \node (V2) [above right=of V1] [rectangle, draw,  fill=gray!20, drop shadow] {$V_2$};
    \node (V3)[below right=of V2] [rectangle, draw,  fill=gray!20, drop shadow] {$V_3$};
    \node (e3) [right =5mm of V3] [scale=0.75, circle, draw, fill=gray!20, drop shadow] {$\varepsilon_3$};
    \node (e2) [right =5mm of V2] [scale=0.75, circle, draw, fill=gray!20, drop shadow] {$\varepsilon_2$};
    \path [->, thick] (V1) edge node [scale=0.75] {$\beta_1$} (V3);
    \path [->, thick] (V3) edge node [scale=0.75] {$\beta_2$} (V2);
    \path [->, thick] (e3) edge node [scale=0.75] {} (V3);
    \path [->, thick] (e2) edge node [scale=0.75] {} (V2);
    \path [<->] (V1) edge [loop above] node [scale=0.75] {$\phi_1$} (V1);
    \path [<->] (V2) edge [loop above, draw=none] node [scale=0.75, opacity=0] {$\phi_2$} (V2);
    \draw [<->] (e3) to [out=20, in=340, looseness=4] node [scale=0.75] {$\psi_3$} (e3);
    \draw [<->] (e2) to [out=20, in=340, looseness=4] node [scale=0.75] {$\psi_2$} (e2);
    \node[above,font=\large\bfseries] at (current bounding box.north) {Model 2A};
\end{tikzpicture}
\end{center}
\end{figure}

We will use the two 3-variable models depicted in Figure \ref{fig:figA}
as our first example (See also Preacher, 2006, p. 228). In Model 1A,
\(V_3\) is regressed on \(V_1\) and \(V_2\), with the covariance among
\(V_1\) and \(V_2\) restricted to zero. Model 2A represents a causal
chain in which \(V_2\) is regressed on \(V_3\), which is in turn
regressed on \(V_1\), yet there is no direct path from \(V_1\) to
\(V_2\). These have the same number of estimated parameters (5) and do
not represent equivalent models, despite the only difference being the
direction of the relationship between \(V_2\) and \(V_3\). The study of
fit propensity is well suited for answering which model has a tendency
to yield better fit. Although these models may seem trivially simple,
the answer to this question is not so easy to see without the additional
work we present below.

\subsubsection{General Procedure and
Code}\label{general-procedure-and-code}

The procedure to study fit propensity that we illustrate here follows
several steps:

\begin{enumerate}
\def\labelenumi{\arabic{enumi}.}
\tightlist
\item
  Definition of the model(s) of interest.
\item
  Generation of \emph{n} random correlation matrices.
\item
  Fitting the models of interest to the \emph{n} random correlation
  matrices.
\item
  Recording information regarding model fit for each model and
  correlation matrix.
\item
  Summaries of model fit using text, graphical displays, and measures of
  effect size (e.g., Komolgorov-Smirnov).
\end{enumerate}

The core custom code used in this paper are included in the
\emph{ockhamSEM} package\footnote{\url{https://github.com/falkcarl/ockhamSEM}}.
Underlying innovations and the methods for generating random correlation
matrices are discussed in the Appendix. The package can be loaded with
the following \emph{R} code snippet:

\singlespace

\begin{verbatim}
library(ockhamSEM)
library(parallel)
\end{verbatim}

\doublespace

\subsubsection{Step 1}\label{step-1}

First use \emph{lavaan} model syntax to define Model 1 and Model 2:

\singlespace

\begin{verbatim}
mod1a <- 'V3 ~ V1 + V2
          V1 ~~ 0*V2'

mod2a <- 'V3 ~ V1
          V2 ~ V3'
\end{verbatim}

\doublespace

Next, two models are fit to data. We expect that this will be the most
typical use of studies of fit propensity for applied researchers - two
or more substantive models are of interest in particular, because of
debates over which is most appropriate for real data. We simply require
a fitted \emph{lavaan} model using data that has the same variable names
(\texttt{V1} through \texttt{V3} in this case) as the above syntax
indicates. Alternatively, we may fit the data to some covariance matrix.
The following uses the latter strategy in creating an identity
matrix:\footnote{Another viable alternative involves generating data
  from the true models for 1A and 2A, such as with \emph{lavaan}'s
  \texttt{simulateData} function.}

\singlespace

\begin{verbatim}
p<-3 # number of variables
temp_mat <- diag(p) # identity matrix

# set row and column names
colnames(temp_mat) <- rownames(temp_mat) <- paste0("V", seq(1, p))
\end{verbatim}

\doublespace

We then fit the two models using the \texttt{sem} function from the
\emph{lavaan} package, though note that any function that returns a
fitted model of class \texttt{lavaan} could be used, such as the
\texttt{cfa}, \texttt{sem}, or \texttt{lavaan} functions:

\singlespace

\begin{verbatim}
mod1a.fit <- sem(mod1a, sample.cov=temp_mat, sample.nobs=500)
mod2a.fit <- sem(mod2a, sample.cov=temp_mat, sample.nobs=500)
\end{verbatim}

\doublespace

At \emph{this} step, any special options regarding estimation can be
passed to \texttt{sem}. Our later code will attempt to use these options
when fitting models for investigating fit propensity. For instance, here
we specify a particular number of observations for this data
(\texttt{sample.nobs=500}), although for many fit indices of interest
this information is inconsequential. One may ask \emph{lavaan} to mimic
a different SEM program, use normal theory or Wishart likelihood, use a
different optimizer, change the iteration limit for estimation, scale
sample covariance matrices by \((N-1)/N\), and so on (see
\texttt{help(lavOptions)}). As long as any of these options are
implemented when defining and fitting initial models, they will be used
when the models are fit to randomly generated correlation matrices.
However, the ability to do so-called robust corrections or use any
estimation approach that requires raw data or a mean structure is not
supported; the available options currently must work for model fitting
when analyzing only a covariance matrix as input.

\subsubsection{Steps 2 through 4}\label{steps-2-through-4}

Generation of random correlation matrices, fitting models to such
matrices, and recording model fit are all accomplished by the
\texttt{run.fitprop} function in the next code snippet:

\singlespace

\begin{verbatim}
res.on <- run.fitprop(mod1a.fit, mod2a.fit, fit.measure="srmr",
                      rmethod="onion",reps=5000,onlypos=TRUE)
\end{verbatim}

\doublespace

The initial arguments to this function are any number of fitted
\emph{lavaan} models, such as \texttt{mod1a.fit} and \texttt{mod2a.fit}.
The remaining arguments must be named and are only required for taking
explicit control over correlation matrix generation and saving of
output. The \texttt{fit.measure} argument accepts a character vector
that indicates what fit indices will be saved. Anything that matches
named output from the \texttt{fitMeasures} command from \emph{lavaan}
can be used. Users are encouraged to run this command on already fitted
models to see what available fit indices are possible (e.g.,
\texttt{fitMeasures(mod1a.fit)}). Here, we save only standardized root
mean square residual, as indicated by \texttt{"srmr"}, which in this
case is equivalent to RMSR since analyzed correlation matrices will
already be standardized. RMSR is the fit index primarily studied by
Preacher (2006) in his work on fit propensity as it provides a sense of
model fit, unadjusted for the number of estimated parameters. We
generate random correlation matrices using the onion method
(\texttt{rmethod="onion"}), requesting 5,000 replications
(\texttt{reps=5000}), and restricting to only positive correlations
(\texttt{onlypos=TRUE}).

The result of the \texttt{run.fitprop} command in the code above is
saved to \texttt{res.on} which is an object of class \texttt{fitprop}
with several options regarding output that will be illustrated shortly.
Before we proceed, suppose we wished to see whether results differ if we
had instead used the MCMC algorithm to generate random correlation
matrices. This latter approach should provide replication of Preacher
(2006), but may be computationally slow. In this case, we may wish to
use parallel processing for faster computations:

\singlespace

\begin{verbatim}
cl <- makeCluster(8)
res.mcmc <- run.fitprop(mod1a.fit,mod2a.fit,fit.measure="srmr",
                        rmethod = "mcmc", reps = 5000, onlypos=TRUE,
                        cluster=cl)
stopCluster(cl)
\end{verbatim}

\doublespace

We create a cluster with 8 processing cores with the
\texttt{makeCluster} command. The result, \texttt{cl}, is then passed to
the \texttt{run.fitprop} function using the \texttt{cluster} argument.
The change in \texttt{rmethod} to \texttt{"mcmc"} will result in use of
the MCMC algorithm for correlation matrix generation. Finally, we shut
down the cluster using the \texttt{stopCluster} command after we obtain
the results.

\subsubsection{Step 5}\label{step-5}

\begin{figure}
\includegraphics{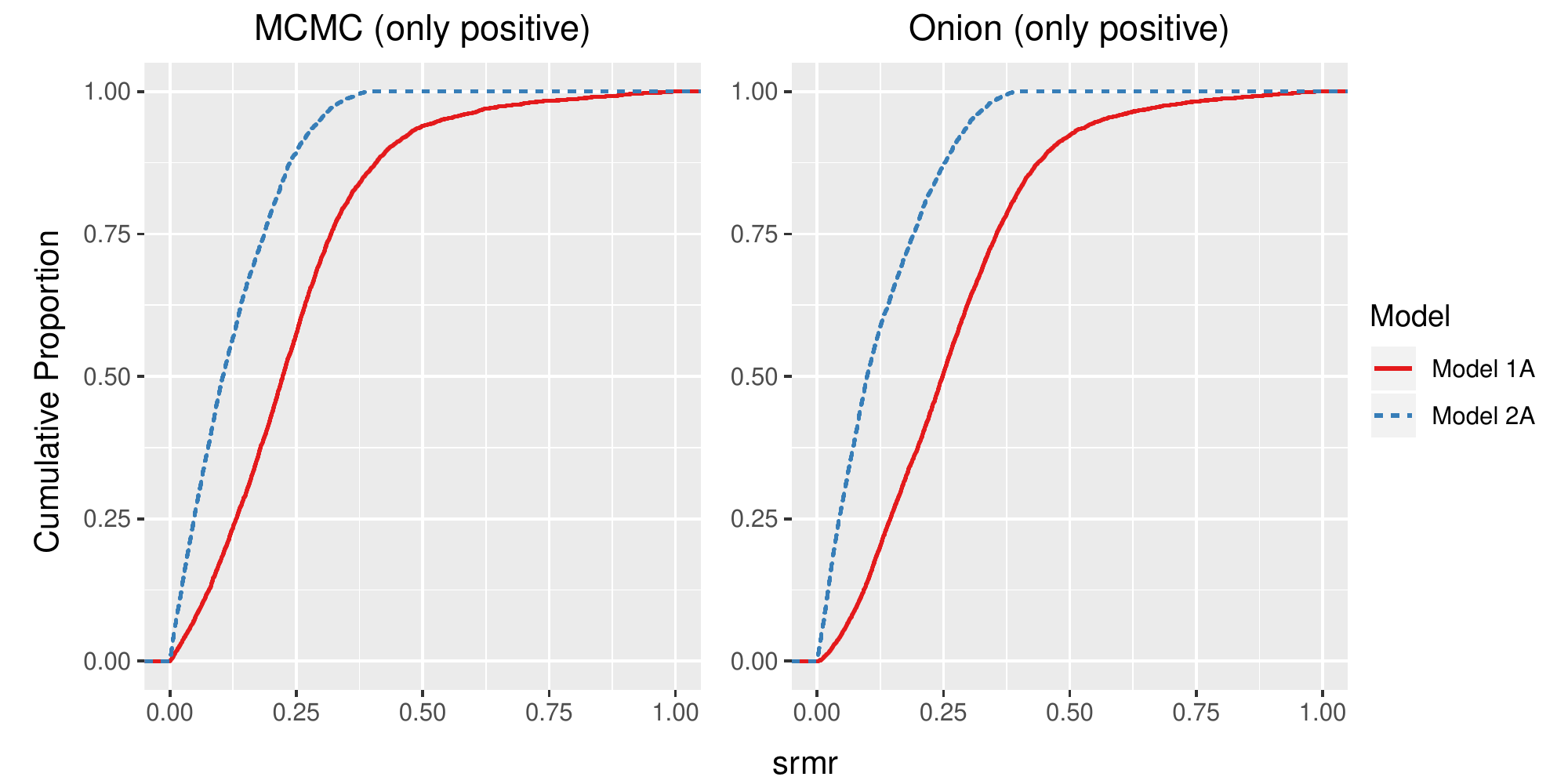} \caption[]{\label{fig:ecdfs1}ECDF plots comparing fit propensity of Models 1A and 2A}\label{fig:unnamed-chunk-10}
\end{figure}

There are multiple different ways to then summarize results. Preacher
(2006) primarily focused on empirical cumulative distribution function
(ECDF) plots that we also illustrate here and will describe shortly. In
particular, both \texttt{res.on} and \texttt{res.mcmc} are objects of
class \texttt{fitprop} for which we have defined a default plot
function. This allows us to simply use the \texttt{plot} command
successively in order to generate ECDF plots (Figure \ref{fig:ecdfs1})
of the requested fit indices and model(s):

\singlespace

\begin{verbatim}
plot(res.on)
plot(res.mcmc)
\end{verbatim}

\doublespace

The argument \texttt{savePlot=TRUE} can also be specified and the result
will be a list of \texttt{ggplot} objects containing ECDF plots
corresponding to each fit index. This feature is useful if, for example,
the user wishes to modify the legend, title, etc., of the resulting plot
or otherwise customize output:\footnote{Which is how we added all of the
  plot titles, combined multiple plots into a single Figure, customized
  axis dimensions and labels, and so on in the present manuscript.} For
convenience, several additional options can be defined, such as to add
custom names (\texttt{mod.lab}) for the two (or more) fitted models and
the color palette used by RColorBrewer (Neuwirth, 2014).

\singlespace

\begin{verbatim}
plot1<-plot(res.on, savePlot=TRUE,
            mod.lab=c("Model 1A","Model 2A"),
            mod.brewer.pal="Set1")
plot2<-plot(res.mcmc, savePlot=TRUE,
            mod.lab=c("Model 1A","Model 2A"),
            mod.brewer.pal="Set1")
\end{verbatim}

\doublespace

To explain ECDFs and Figure \ref{fig:ecdfs1}, suppose we collect all
RMSR estimates for the 5,000 fitted models for Model 1A. We then sort
these 5,000 estimates in order from lowest to highest. Next, we count
the number of RMSR estimates at or below a particular value. For
example, ``what proportion of fitted models have an RMSR value of .25 or
lower? .5 or lower?'' Each curve in Figure \ref{fig:ecdfs1} displays the
answer to this question for each model separately and at many values of
RMSR along the x-axis such that the lines appear to be continuous. For
example, Model 2A had approximately 75\% (or .75 as a proportion) of
models that had an RMSR (or srmr) of .25 or better when correlation
matrices were generated using the MCMC algorithm (see where .25 on the
x-axis intersects with the dotted blue line on the left-hand panel).
Model 1A had a smaller proportion (around .57 or so) of cases with an
RMSR of .25 or better. This implies that the higher curve (for Model 2A)
indicates better fit for more models, and therefore more fit propensity
when examining RMSR.

For the most part, the results displayed here replicate those of
Preacher (2006): Model 2A appears to have more fit propensity in that
there is a higher proportion of RMSR values that are relatively small,
and this result tends to hold for both MCMC and Onion methods. Had
either correlation matrix generating method allowed for both positive
and negative correlations, the basic pattern in fit propensity regarding
Models 1A and 2A would have been similar and the interested reader is
encouraged to verify this assertion.

Default \texttt{print} and \texttt{summary} methods are also available
for \texttt{fitprop} objects. \texttt{summary} will provide some
diagnostic information regarding whether any non-convergent models were
encountered, selected quantiles of the resulting fit statistics, and
effect sizes to help quantify the differences in fit between estimated
models. For example, instead of eyeballing Figure \ref{fig:ecdfs1}, we
can ask for the value of \texttt{srmr} that corresponds to a cumulative
proportion of .25, .5, and .75 for both models using the following:

\singlespace

\begin{verbatim}
summary(res.mcmc, probs=c(.25,.5,.75))
\end{verbatim}

\begin{verbatim}
## 
##  Quantiles for each model and fit measure:
## 
##  Model  1 
##      srmr
## 25% 0.134
## 50% 0.229
## 75% 0.330
## 
##  Model  2 
##      srmr
## 25% 0.047
## 50% 0.105
## 75% 0.187
## 
##  Information about replications for each model and fit measure:
## 
##  Model  1 
## 
## Mean across replications
##  srmr 
## 0.298 
## 
## Median across replications
##  srmr 
## 0.229 
## 
## Number of finite values
## srmr 
## 5000 
## 
## Number of NA values
## srmr 
##    0 
## 
##  Model  2 
## 
## Mean across replications
##  srmr 
## 0.123 
## 
## Median across replications
##  srmr 
## 0.105 
## 
## Number of finite values
## srmr 
## 5000 
## 
## Number of NA values
## srmr 
##    0 
## 
##  Effect Sizes for Differences in Model Fit:
## 
##   srmr 
## 
##  Model 1 vs. Model 2 
##    Cohen's d:           0.607 
##    Cliff's delta:       0.512 
##    Komolgorov Smirnov:  0.374
\end{verbatim}

\doublespace

These results tell a similar story to the graphical summaries provided.
For example, 50\% of Model 1A results had an RSMR of .229 or lower,
whereas 50\% of Model 2A had an RSMR of .105 or lower. Alternatively,
one may examine the mean or median values for RMSR. We also see that
there were apparently no models where non-convergence was a problem,
since there are no NA values for any \texttt{srmr} estimates.

At the end of the output, differences between all estimated models and
all recorded fit indices are quantified using three effect sizes:
Cohen's \(d\) (Cohen, 1988), Cliff's delta (Cliff, 1996), and a
Komolgorov-Smirnov coefficient (K-S). Although additional effect sizes
could be easily added, we initially chose these three for several
reasons. First, Cohen's \(d\) is likely familiar to many researchers in
the social sciences. Here, we see that a value of .61 is observed,
which is typically considered as between a ``medium'' and 
``large'' effect size in research settings. Here it is indicating that Model 1A's distribution for RMSR is on average .61 SD higher than Model 2A. Cohen's \(d\) has a nice
conceptual interpretation as the number of standard deviation units that
separate the RMSR distributions for Models 1A and 2A. Since Cohen's
\(d\) is typically not considered a robust effect size measure (e.g.,
Wilcox, 2012), we included Cliff's delta, which is robust to outliers
and skewness, and K-S as additionally sensitive to variability across
two distributions. These may be more unfamiliar to researchers.

Conceptually, Cliff's delta is a difference between two probabilities: 1) the probability that a value in the first distribution is greater than that in the second; and 2) the probability that a value in the second distribution is greater than that in the first. It is computed in part by comparing each observation from one distribution versus all observations in a second distribution. Cliff's delta ranges between -1 and 1, with values
close to zero indicating no difference between two distributions. The K-S coefficient is the maximum difference between two ECDFs over all cumulative probabilities of some measure (in our case, a given fit index). Therefore, as ECDFs also are bound between 0 and 1, K-S ranges between 0 and 1, with 1 indicating a larger discrepancy between
two distributions.\footnote{Although K-S could also yield a \(p\)-value,
  this value would largely be dependent on the number of replications
  chosen for the fit propensity analysis.} To our knowledge, there are no accepted guidelines for Cliff's delta or the K-S statistic equivalent to "small", "medium" and "large" for Cohen's d, in part because the statistics are readily interpretable. For example, in the output above, the difference in probabilities is 51.2\%  with values tending to be higher in Model 1,  and the largest discrepancy in ECDFs is .374.

\subsection{Example B: Simplex model versus factor
model}\label{example-b-simplex-model-versus-factor-model}

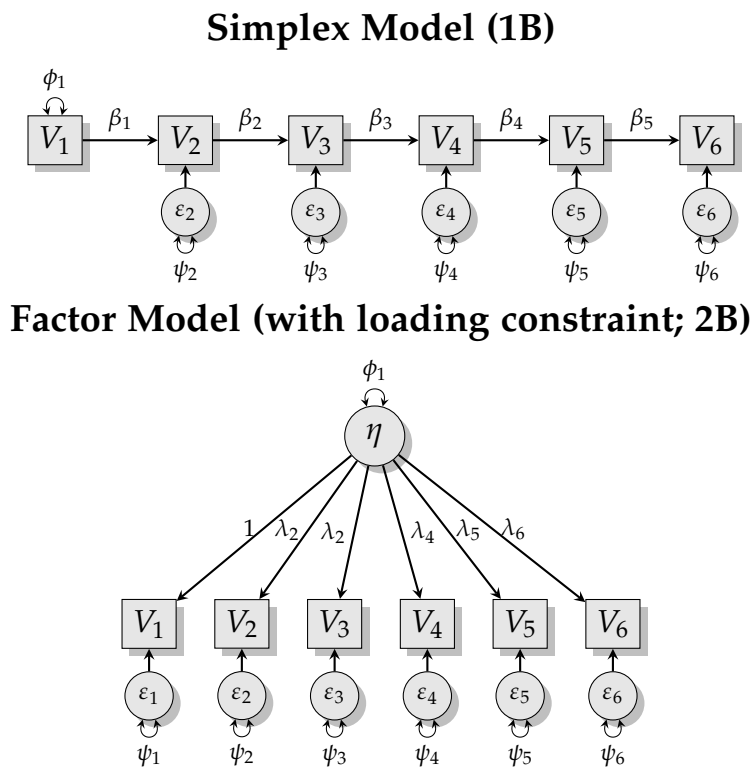
\begin{figure}[h]
\begin{center}
\caption{Simplex and constrained factor analysis model.}  \vskip10pt
\label{fig:figB}
\begin{tikzpicture}[auto, >=stealth,
    error/.style={edge from parent/.style={draw, <-}},
    level 1/.style={level distance=5mm}]
    \node (V1) [rectangle, draw,  fill=gray!20, drop shadow] {$V_1$};
    \node (V2) [right=of V1][rectangle, draw, fill=gray!20, drop shadow] {$V_2$};
  \node (V3) [right=of V2][rectangle, draw, fill=gray!20, drop shadow] {$V_3$};
    \node (V4) [right=of V3][rectangle, draw, fill=gray!20, drop shadow] {$V_4$};
    \node (V5) [right=of V4][rectangle, draw, fill=gray!20, drop shadow] {$V_5$};
    \node (V6) [right=of V5][rectangle, draw, fill=gray!20, drop shadow] {$V_6$};
    \node (e2) [below=3mm of V2][scale=0.75, circle, draw, fill=gray!20, drop shadow] {$\varepsilon_2$};
    \node (e3) [below=3mm of V3][scale=0.75, circle, draw, fill=gray!20, drop shadow] {$\varepsilon_3$};
    \node (e4) [below=3mm of V4][scale=0.75, circle, draw, fill=gray!20, drop shadow] {$\varepsilon_4$};
    \node (e5) [below=3mm of V5][scale=0.75, circle, draw, fill=gray!20, drop shadow] {$\varepsilon_5$};
    \node (e6) [below=3mm of V6][scale=0.75, circle, draw, fill=gray!20, drop shadow] {$\varepsilon_6$};
    \path [<->] (V1) edge [loop above] node [scale=0.75] {$\phi_1$} (V1);
    \path [->, thick] (V1) edge node [scale=0.75] {$\beta_1$} (V2);
    \path [->, thick] (V2) edge node [scale=0.75] {$\beta_2$} (V3);
    \path [->, thick] (V3) edge node [scale=0.75] {$\beta_3$} (V4);
    \path [->, thick] (V4) edge node [scale=0.75] {$\beta_4$} (V5);
    \path [->, thick] (V5) edge node [scale=0.75] {$\beta_5$} (V6);
    \path [->, thick] (e2) edge node [scale=0.75] {} (V2);
    \path [->, thick] (e3) edge node [scale=0.75] {} (V3);
    \path [->, thick] (e4) edge node [scale=0.75] {} (V4);
    \path [->, thick] (e5) edge node [scale=0.75] {} (V5);
    \path [->, thick] (e6) edge node [scale=0.75] {} (V6);
    \path [<->] (e2) edge [loop below] node [scale=0.75] {$\psi_2$} (e2);
    \path [<->] (e3) edge [loop below] node [scale=0.75] {$\psi_3$} (e3);
    \path [<->] (e4) edge [loop below] node [scale=0.75] {$\psi_4$} (e4);
    \path [<->] (e5) edge [loop below] node [scale=0.75] {$\psi_5$} (e5);
    \path [<->] (e6) edge [loop below] node [scale=0.75] {$\psi_6$} (e6);
    \node[above,font=\large\bfseries] at (current bounding box.north) {Simplex Model (1B)};
\end{tikzpicture}

\begin{tikzpicture}[>=stealth]
\node[ov] (v1) at (0,0) {$V_1$};
\node[ov] (v2) [right=5mm of v1]  {$V_2$};
\node[ov] (v3) [right=5mm of v2]  {$V_3$};
\node[ov] (v4) [right=5mm of v3]  {$V_4$};
\node[ov] (v5) [right=5mm of v4]  {$V_5$};
\node[ov] (v6) [right=5mm of v5]  {$V_6$};
\node[lv] (F) at (3,2.5) {$\eta$};
\node (e1) [below=3mm of v1][scale=0.75, circle, draw, fill=gray!20, drop shadow] {$\varepsilon_1$};
\node (e2) [below=3mm of v2][scale=0.75, circle, draw, fill=gray!20, drop shadow] {$\varepsilon_2$};
\node (e3) [below=3mm of v3][scale=0.75, circle, draw, fill=gray!20, drop shadow] {$\varepsilon_3$};
\node (e4) [below=3mm of v4][scale=0.75, circle, draw, fill=gray!20, drop shadow] {$\varepsilon_4$};
\node (e5) [below=3mm of v5][scale=0.75, circle, draw, fill=gray!20, drop shadow] {$\varepsilon_5$};
\node (e6) [below=3mm of v6][scale=0.75, circle, draw, fill=gray!20, drop shadow] {$\varepsilon_6$};
\path[->, thick] (F) edge node[scale=0.75, left] {1} (v1)
(F) edge node[scale=0.75, left] {$\lambda_2$} (v2)
(F) edge node[scale=0.75, left] {$\lambda_2$} (v3)
(F) edge node[scale=0.75, right] {$\lambda_4$} (v4)
(F) edge node[scale=0.75, right] {$\lambda_5$} (v5)
(F) edge node[scale=0.75, right] {$\lambda_6$} (v6);
\path [<->] (F) edge [loop above] node [scale=0.75] {$\phi_1$} (F);
\path [<->] (e1) edge [loop below] node [scale=0.75] {$\psi_1$} (e1);
\path [<->] (e2) edge [loop below] node [scale=0.75] {$\psi_2$} (e2);
\path [<->] (e3) edge [loop below] node [scale=0.75] {$\psi_3$} (e3);
\path [<->] (e4) edge [loop below] node [scale=0.75] {$\psi_4$} (e4);
\path [<->] (e5) edge [loop below] node [scale=0.75] {$\psi_5$} (e5);
\path [<->] (e6) edge [loop below] node [scale=0.75] {$\psi_6$} (e6);
\path [->, thick] (e1) edge node [scale=0.75] {} (v1);
\path [->, thick] (e2) edge node [scale=0.75] {} (v2);
\path [->, thick] (e3) edge node [scale=0.75] {} (v3);
\path [->, thick] (e4) edge node [scale=0.75] {} (v4);
\path [->, thick] (e5) edge node [scale=0.75] {} (v5);
\path [->, thick] (e6) edge node [scale=0.75] {} (v6);
\node[above,font=\large\bfseries] at (current bounding box.north) {Factor Model (with loading constraint; 2B)};
\end{tikzpicture}
\end{center}
\end{figure}

Next, we extend an example given by Preacher (2006) in which we compare
a simplex model (1B) with a single factor model with a loading equality
constraint for the second and third loadings (2B; Figure
\ref{fig:figB}).

Although it seems unlikely that researchers would consider these two
alternative models for the same dataset, they have the same degrees of
freedom and will yield different fit propensity. Furthermore, these
examples are useful for demonstrating the impact of a restriction on the
data space. In particular, we compared the fit propensity using
\texttt{srmr} for these two models by crossing two conditions:
Correlation matrix generation (MCMC versus Onion) and positivity of
correlations (all positive versus positive and negative). The option for
obtaining both positive and negative correlations can be achieved by
setting \texttt{onlypos=FALSE} when using the \texttt{run.fitprop}
command.\footnote{R code for this and for all following examples appears
  in Supplementary Materials.}

\blandscape

\begin{figure}
\includegraphics{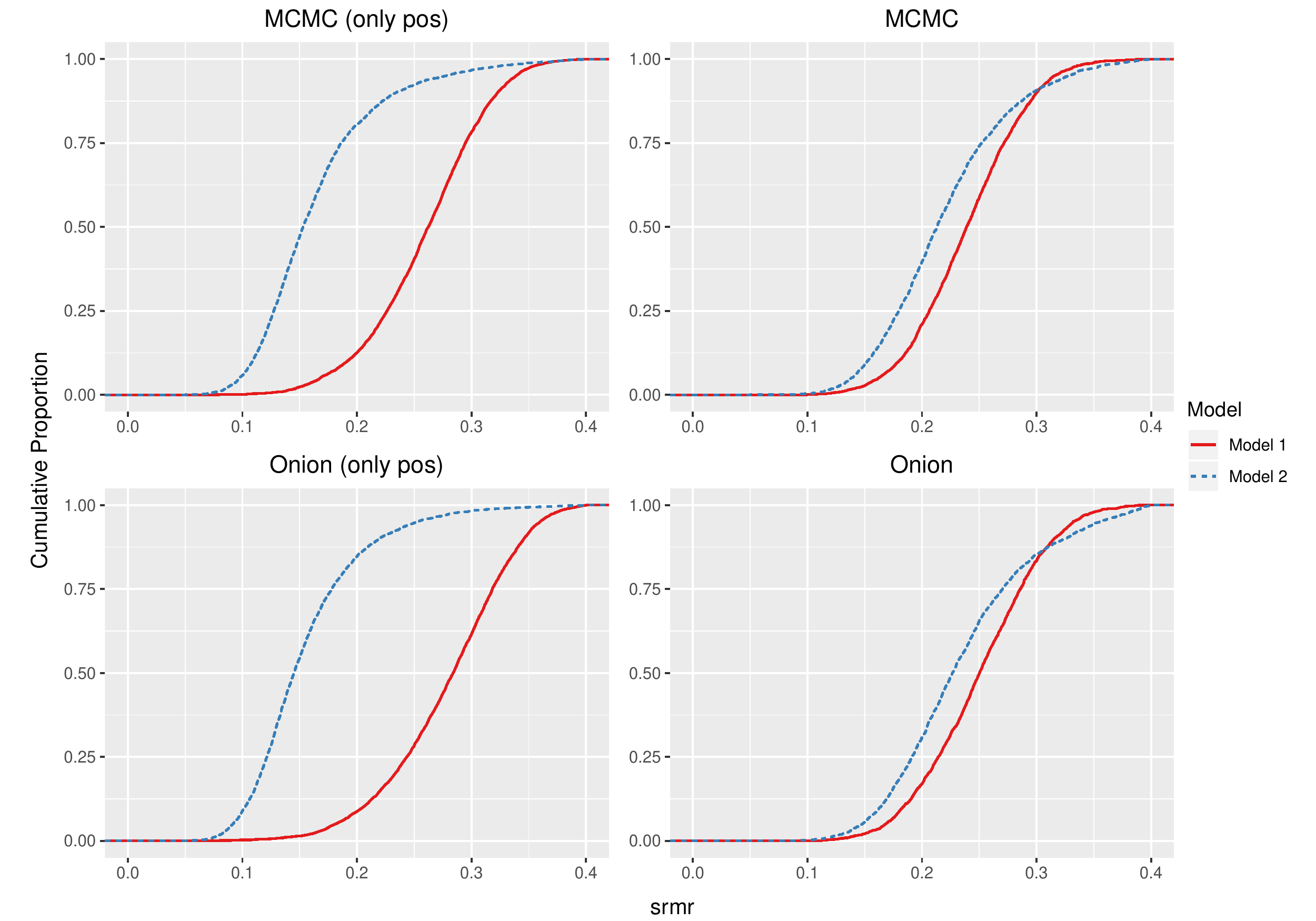} \caption[]{\label{fig:ecdfs2}ECDF plots comparing fit propensity of Models 1B and 2B}\label{fig:unnamed-chunk-16}
\end{figure}

\elandscape

Here, we see that the restriction of the data space to only positive
correlations makes a substantial difference, yet also note that Preacher
(2006) only examined positive correlations. The factor model (2B)
generally has more fit propensity than the simplex model (1B) if only
positive correlations are considered as shown in the two plots on the
left-hand side of Figure \ref{fig:ecdfs2}. However, the fit propensity
for the factor model is lower if we consider the possibility that
the data space may include both positive and negative correlations as
shown on the right-hand side of Figure \ref{fig:ecdfs2}. In retrospect,
this makes intuitive sense given that the second and third loadings of
the factor model are constrained equal. At minimum, such a restriction
would likely only make sense if the correlations among these two items
have the same sign. For example, if it were the case that item 2 tended
to have negative correlations with other items, but item 3 had positive
correlations, the two loadings would seem to have opposite signs. Model
fit would then deteriorate due in part to constraining these loading
estimates to be equal.

We also note that substantial estimation problems were encountered when
the correlation matrices included both positive and negative values. In
particular, 2462 models failed to converge for the factor model when
negative correlations were allowed for the Onion method, but 0 and 16
failed to converge for the MCMC and Onion (only positive) conditions,
respectively. Such convergence failures are indicated as \texttt{NA}
values for particular fit indices. Such information regarding the number
of valid replications is available via the \texttt{summary} command.

The above plots and summary information are based on replications where
\emph{both} models converged as this is the default behavior. We can,
however, change output so that results are based on the available number
of replications for either model, regardless of whether the replications
are the same by setting \texttt{samereps=FALSE} when using the
\texttt{plot} command (For example,
\texttt{summary(res.mcmc,\ samereps=FALSE)}). Thus, the quantiles and
plots for Model 1B could be based on a slightly different subset of
random correlation matrices than Model 2B. However, this does not appear
to substantially change the resulting plots or summary information, and
the interested reader is encouraged to verify this observation.

\subsection{Models for the Rosenberg Self-Esteem
Scale}\label{models-for-the-rosenberg-self-esteem-scale}

In our final example, we return to the debate regarding an appropriate
model for the RSES. In particular, we investigated the fit propensity of
the four models considered in Table \ref{tbl:RSESfit}. For reference, we
label these as follows: Correlated residuals (Model 1C), bifactor (Model
2C), bifactor with correlated method factors (Model 3C), and a single
factor model (Model 4C). Due to more models and the possibility of
nonconvergence, we increased the number of replications to 10,000. As
the RSES items typically intercorrelate positively (after
reverse-coding), we used the onion method with only positive
correlations. For fit indices, we additionally demonstrate the
performance of TLI, CFI, RMSEA, and RMSR. The model chi-square is also
saved to later check model nesting.

Since information is saved from four fit indices (TLI, RMSR, CFI, RMSEA)
and the model chi-square test of fit, we illustrate two additional
options for aiding in generating plots. First, we can either obtain
plots for all fit indices (the default behavior), or request plots just
for a particular fit index by passing the \texttt{whichfit} argument to
the \texttt{plot} function. \texttt{whichfit} accepts a character vector
in the same fashion as \texttt{fit.measure} (resulting in plots for
selected fit indices being displayed successively), or may be left at
its default setting (successively showing plots for all saved fit
indices).

Second, an additional argument can be changed to aid in interpreting TLI
and CFI. Note that lower values of RMSR typically indicate better fit,
but \emph{higher} values of TLI and CFI correspond to better fit. The
\texttt{lower.tail} argument accepts a logical vector of the same length
as the number of stored fit indices (i.e., the same length as
\texttt{fit.measure}). By default, the vector sets all elements to
\texttt{TRUE}. Setting the corresponding element to \texttt{FALSE} for
CFI ensures that the ECDF plots start with higher numbers on the
left-hand side of the x-axis, making interpretation of such plots
visually similar to those for RMSR. In the following code snippet,
assuming results are stored in \texttt{res.rses} we only plot results
for TLI, CFI, RMSEA and RMSR, and we make the above corresponding
changes for the ECDF plot for TLI and CFI.

\begin{verbatim}
plot(res.rses, whichfit=c("tli","cfi","rmsea","srmr"),
     lower.tail=c(FALSE,FALSE,TRUE,TRUE))
\end{verbatim}

\begin{figure}
\includegraphics{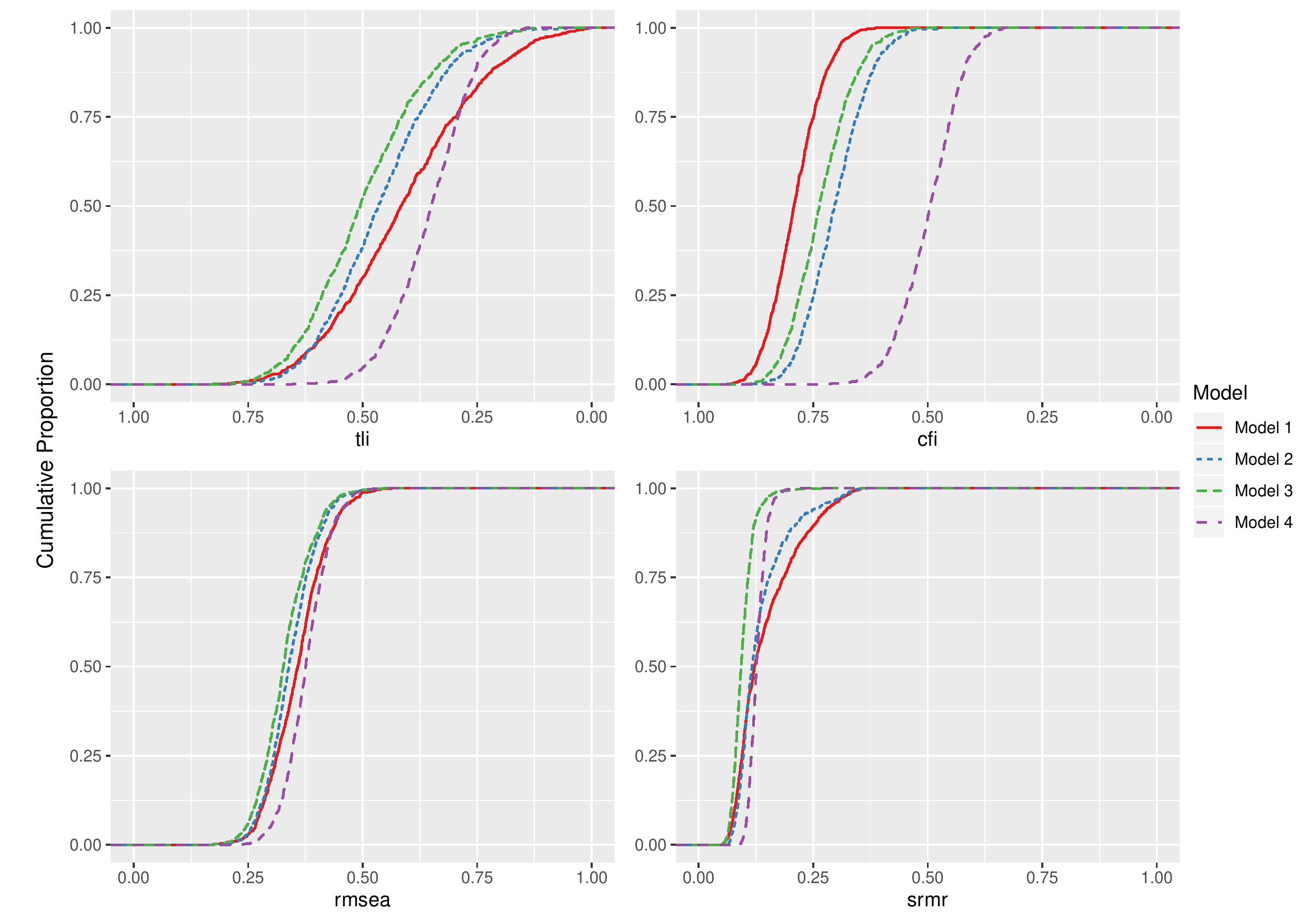} \caption[]{\label{fig:ecdfs4}ECDF plots comparing fit propensity of Models 1C, 2C, 3C, and 4C}\label{fig:unnamed-chunk-23}
\end{figure}

We next examine each fit index separately (Figure \ref{fig:ecdfs4}). CFI
is perhaps the easiest to interpret, yet also raises the most concerns
about its use as a measure of model fit. In particular, very clearly the
four models are ordered in terms of the number of estimated parameters,
with models with fewer degrees of freedom (more estimated parameters)
having more fit propensity according to CFI. This is intuitive to the
extent that more parameters leads to better fit. However, this is
concerning since the data are random and CFI should include an
adjustment for parsimony. That is, the adjustment included in CFI does
not appear to equate the fit of the resulting models. Even for some
nontrivial percentage of replications, CFI even exceeds .80 or so (e.g.,
CFI averages around .79 for the correlated residual model). This large
percentage of models that have good fit for random data should be
concerning to researchers using CFI.

TLI and RMSEA tend to have similar patterns of results to each other.
Almost no replications are in the range of what is usually considered
acceptable fit, and models are ordered in a non-intuitive way: The
bifactor model with correlated method factors had the highest fit
propensity, followed by the bifactor model, correlated residual model,
and finally the one factor model. Thus, the ordering of fit propensity
is not in line with the \emph{df} or number of estimated parameters in
that the correlated residual model did not have the highest fit propensity.

Differences among the models are also more difficult to detect visually.
Additionally, effect sizes for the differences across these
distributions also tends to be smaller. For example, K-S yields values
between .11 and .40 for all pair-wise differences for RMSEA. For RMSR,
the bifactor model with correlated method factors yielded the highest fit
propensity with the other models mixed. For instance, the difference
between the bifactor model and the correlated residual is negligible
(Cliff's delta = .05, K-S = .11). However, we may also raise some
concerns about the utility of RMSR for these studied models. For
instance, many RMSR estimates were below .1, with a full 60\% of the
correlated bifactor models below this threshold.

Euler plots can further aid in helping visualize relative fit of the
models for the same randomly generated correlation matrices and are
similar to Venn diagrams. These are inspired by the ``ameoba'' plots of
Bonifay \& Cai (2017). In Euler plots, the area of each circle (or
ellipse) is proportional to the number of cases that meet some criteria.
Overlap among circles indicates overlap in the sets of correlation
matrices that meet this criteria.\footnote{Code using \texttt{nVennR}
  for generating plots is also currently in development.}

Take for example, TLI. The ECDF plot seems to indicate that the
correlated bifactor model had about 50\% of replications with a TLI of
.5 or better, whereas the single factor model had only about 8\% of
replications with a TLI of .5 or better. But, does the correlated
bifactor model \emph{always} fit better according to TLI? In other
words, consider the entire set of correlation matrices that had TLI of
.5 or better for the correlated bifactor model and the set of
correlation matrices that had TLI of .5 or better for the single factor
model. Are there some replications where the single factor model had a
TLI better than .5, but the correlated bifactor model did not? This
information regarding overlapping sets of replications that meet this
TLI criterion is depicted in a Euler plot:

\begin{verbatim}
plot(res.rses,type="euler",whichfit="tli",
     whichmod=c(3,4),cutoff=.5,lower.tail=FALSE)
\end{verbatim}

\begin{figure}
\includegraphics{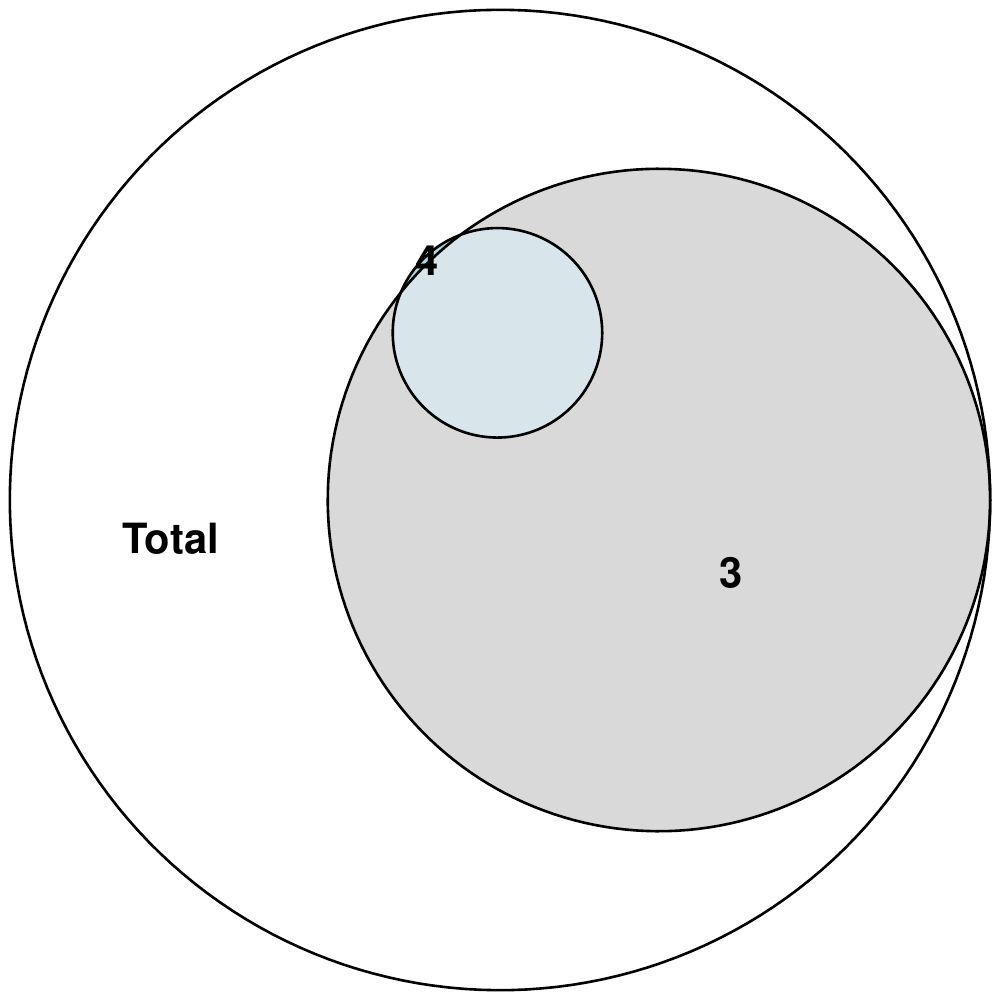} \caption[]{\label{fig:euler31}Euler plot for TLI at cutoff=.5}\label{fig:unnamed-chunk-24}
\end{figure}

Figure \ref{fig:euler31} was generated with the above code. Larger
ellipses are conceptually similar to occupation of a wider area of the
data space, or more fit propensity, and directly represent the
proportion of correlation matrices for each model where TLI was better
than .5. Here we see that the correlated bifactor model takes up the
most space relative to the single factor model, indicating that it had a
higher proportion of cases with TLI of .5 or better. If one model
\emph{always} fits better than another, we would expect that the worse
fitting model would be completely contained within the better fitting
model's ellipse. Such a case would indicate that the entire set of
replications that met the \(TLI\ge .5\) criterion for the lesser fitting
model was also met by the better fitting model. That each model has some
part of the ellipse that is not covered by the other suggests that there
are some replications where each model fits better than the others. Here
we see that the single factor model is \emph{not} completely contained
within the correlated bifactor's space. This means that there are some
correlation matrices for which the single factor model has TLI better
than .5, but the bifactor model does \emph{not} have TLI better than .5.
In other words, there are some correlation matrices for which the single
factor model has better TLI than the correlated bifactor model.

\begin{figure}
\includegraphics{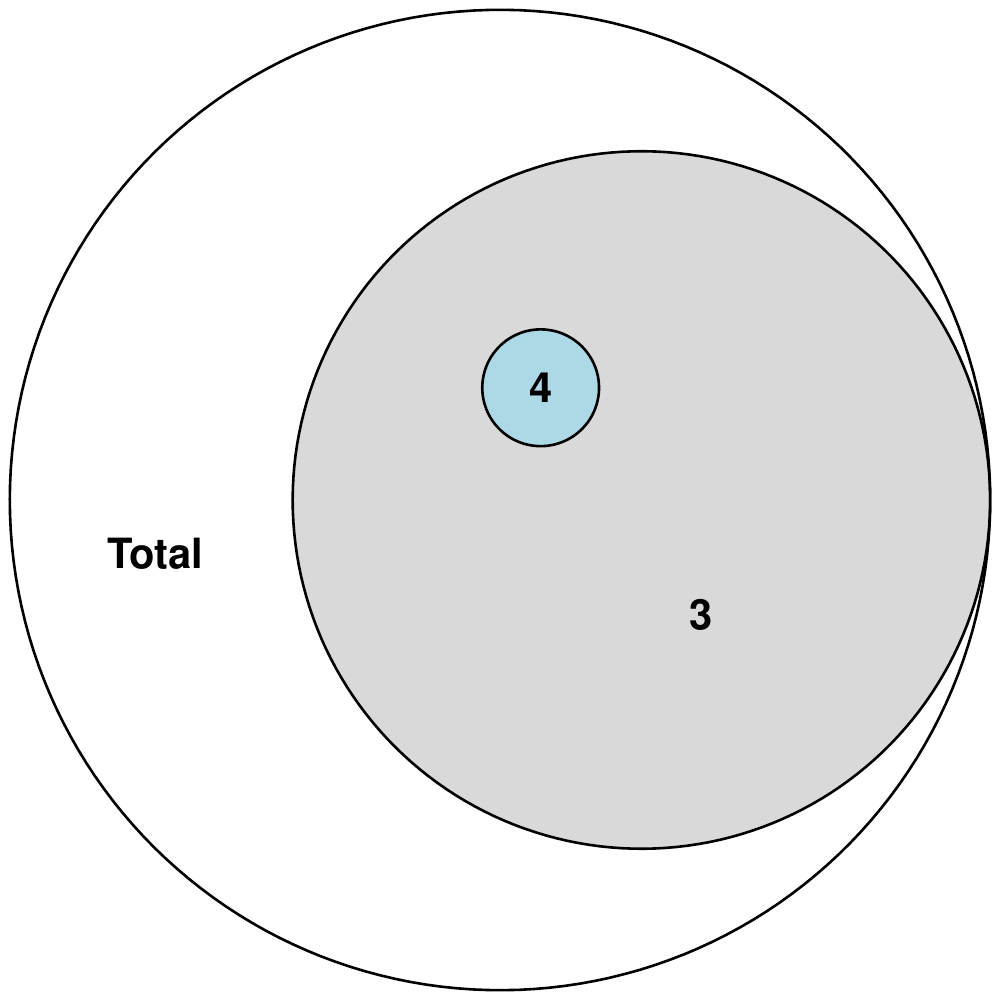} \caption[]{\label{fig:euler32}Euler plot for chi-square test at cutoff=3000}\label{fig:unnamed-chunk-25}
\end{figure}

Although one might be under the impression that models with more factors
should always fit better than a single Factor model, this result
typically applies to the chi-square test of fit and to the case of
nested models. We can, however, directly check to see if the 1-Factor
model is nested with a correlated bifactor model by also examining Euler
plots for the chi-square test of model fit (Figure \ref{fig:euler32}).
Here, we used a cutoff value of \(\chi^2 = 3,000\). The single factor
model has no unique set that is not completely encompassed by the
correlated bifactor model. In other words, there do not appear to be any
single factor models that are better than \(\chi^2=3,000\) but for which
a correlated bifactor model is not as good. These results are what we
would expect with a unidimensional model being nested with the
correlated bifactor model.\footnote{As a small aside, proper examination
  of model nesting is contingent upon good starting values and
  successful estimation. For example, \emph{lavaan}'s defaults for
  starting values can sometimes converge on a local optimum and thereby
  yield a higher chi-square value even for a more general model. Should
  the researcher wish to examine particular replications, randomly
  generated correlation matrices and fitted models can be saved as lists
  to the \texttt{res.rses} object.}

Finally, SIC as reported in Table \ref{tbl:RSESfit} has been suggested
by others (Bonifay \& Cai, 2017; see also Hansen \& Yu, 2001; Preacher,
2006) as a computationally feasible index that is similar to AIC and
BIC, but may make a more fine-grained adjustment than that based on the
number of estimated parameters.\footnote{Like AIC and BIC, SIC is
  computed in part from the negative log-likelihood, but makes an
  adjustment based on the log of the determinant of the information
  matrix for the item parameters. Thus, more parameters may lead to a
  greater adjustment, but to the extent that parameter estimates are
  asymptotically correlated the adjustment may be less.} In this
particular case, SIC suggests the same ordering of model fit, and
therefore the same selected model, as these other more traditional
indices. While \texttt{semTools} (Jorgensen, Pornprasertmanit,
Schoemann, \& Rosseel, 2019) now provides computation of SIC from fitted
\emph{lavaan} models, we note that SIC does not immediately provide us
with the useful information regarding fit propensity and other fit
indices that may be of interest. The method of studying fit propensity
that we have discussed thus far also allows restriction of the data
space, which is also not possible with SIC.

\doublespace

\section{Discussion and Conclusion}

Structural Equation Models (SEMs) are a flexible method for representing
complex theories, a useful tool in moving toward a formal theoretical
approach to the psychological sciences (Muthukrishna \& Henrich, 2019).
But in evaluating competing theories instantiated as different SEMs,
researchers need to consider not only the fit of their models to the
data, but the parsimony of that fit. Fit propensity is therefore an
important consideration in evaluating the relative merit of competing
theories.

We introduce and investigate fit propensity using several examples,
including investigation of the Rosenberg Self Esteem Scale. These
investigations reveal important caveats in evaluating model fit and
multimodel inference. First, they make clear that model fit alone is not
sufficient to select between competing models--fit propensity also needs
to be considered. Second, they reveal that fit indices, particularly
RMSR and CFI, evaluated in isolation can be misleading. For example, in
the RSES example, CFI was found to have decent for fit for a large percentage of random
correlation matrices - as high as 80\% in some cases. Finally, this
investigation provides a replication of Preacher's (2006) original
findings regarding fit propensity, extending these beyond positive
correlation matrices. These examples also illustrate the usefulness of
the \emph{ockhamSEM} package. The package is intended to support
researchers in considering the fit propensity of their SEMs, but also
support applied researchers and methodologists in further investigating
fit propensity itself and refining methods for evaluating fit
propensity.

Several challenges remain that we hope will be addressed by methodological researchers. One practical avenue for refinement is computational complexity. In
Preacher's (2006) original code generating random correlation matrices
was a significant computational bottleneck. We solved this problem by
using the onion method (Joe, 2006; Lewandowski et al., 2009). Model
fitting, however, remains computationally intensive, and this problem
increases with more variables. Models with more variables may also
encounter more estimation problems. Some avenues for solving this
problem include integrating our R code with a structural equation
modeling program that might fit models faster (e.g., Bentler, 2006; L.
K. Muthèn \& Muthèn, 1998--2010; Neale et al., 2016). OpenMx may be the
most promising alternative as it also allows access to optimizers that
better overcome local optimum or automate attempts at different starting
values.

Another important area of further research is constraints on the space
of data. Example B reveals how restricting the data space to only
positive correlations versus allowing positive and negative correlations
can affect fit propensity. Ideally the data space could be restricted
based on the universe of possible data for a particular research
question. In future research, we hope to investigate the generation of
random correlation matrices from specific data spaces for a fixed number
of factors (where the number of factors is less than the number of
items), or from the entire data space under a particular theoretical
model of interest. Lai et al. (2017) provided an initial attempt at
something similar, but their approach is not guaranteed to generate
uniformly from the entire data space under any given model. Additional
tuning parameters are available under Lewandowski et al. (2009), but are
challenging to translate for the applied researcher.

As our illustrations reveal, fit propensity and number of parameters should be interpreted independently when assessing parsimony. Researchers can only attribute the difference in fit propensity to model specification, or make general statements about types of model specifications and their relative fit propensity when number of parameters are the same. For instance, Examples A and B illustrate cases where fit propensity is different even though competing models have the same number of estimated parameters. This will of course not always be the case in the real world, for example, when assessing two theoretically derived model specifications, as is the case with the final RSES example. Assessing fit propensity gives us another tool for claiming parsimony, even making the case for a model not only having fewer parameters, but also higher fit propensity (Bonifay \& Cai, 2017; Pearl \& Verma, 1995). For example, for some indices, the most unrestricted RSES model did not have the highest fit propensity. It is also not easy to separate the effects on fit propensity due to the number of estimated parameters from the configural (or functional) form of the model. For instance, even though TLI, CFI, and RMSEA make some adjustment for \textit{df}, they yielded different conclusions in the RSES example.

Beyond this tutorial, some researchers may desire additional reading on the background literature that gave rise to fit propensity and its relation to other fields. To briefly elaborate, the minimum description length (MDL; Rissanen, 1978) principle is one key concept in information theoretic perspectives regarding model fit (e.g., see Bonifay, 2015) and is often used in cognitive, computer science, and machine learning.  MDL considers the data to be composed of both noisy and systematic components. The goal is often to find a way to encode the sytematic part in such a way that both systematic (i.e., model) and noisy parts (i.e., remaining error) can be most concisely encoded. Thus, MDL formalizes Occam's razor and the tradeoff between a complicated model (i.e., not concisely describing the systematic part) and poor model fit (i.e., too much remaining error to describe). Importantly, the MDL perspective does not require that there is a ``true'' population model. However, it is often used for inductive inference as the model that allows description of the data in the most concise manner is often one that allows us to learn much from the data (it is useful) and should be considered the best model. Researchers wishing to have a better understanding of these concepts or indices more directly derived from MDL (such as normalized maximum likelihood) are referred to resources that we believe are accessible in part by focusing on such concepts within SEM and IRT frameworks (Bonifay, 2015; Bonifay \& Cai, 2017; Preacher, 2005, 2006).

While there are practical considerations in evaluating fit propensity
for model selection, it is clear that fit propensity cannot simply be
ignored--parsimony is a crucial aspect of evaluating theories. That a
model better fits the right data is insufficient evidence if the model
also fits the wrong data. We hope to motivate researchers to always
consider parsimony and fit propensity and provide them with the tools to
do so.

\clearpage

\section{References}

\setlength{\parindent}{-24pt} \setlength{\leftskip}{24pt}
\setlength{\parskip}{0pt} \noindent

\hypertarget{refs}{}
\hypertarget{ref-Bentler1990}{}
Bentler, P. M. (1990). Comparative fit indexes in structural models.
\emph{Psychological Bulletin}, \emph{107}(2), 238--246.
\url{http://doi.org/10.1037/0033-2909.107.2.238}

\hypertarget{ref-Bentler:2006:B}{}
Bentler, P. M. (2006). \emph{EQS 6 structural equations program manual}.
Encino, CA: Multivariate Software Inc.

\hypertarget{ref-Bentler2010}{}
Bentler, P. M., \& Satorra, A. (2010). Testing model nesting and
equivalence. \emph{Psychological Methods}, \emph{15}(2), 111--123.
\url{http://doi.org/10.1037/a0019625}

\hypertarget{ref-Bollen:2013:IC}{}
Bollen, K. A., \& Pearl, J. (2013). Eight myths about causality and
structural equation models. In S. L. Morgan (Ed.), \emph{Handbook of
causal analysis for social research} (pp. 301--328). Dordrecht: Springer
Netherlands. \url{http://doi.org/10.1007/978-94-007-6094-3_15}

\hypertarget{ref-Bonifay:PhD}{}
Bonifay, W. E. (2015). \emph{An integrative framework of model
evaluation} (PhD thesis). Department of Psychology, University of
California, Los Angeles.

\hypertarget{ref-Bonifay2017}{}
Bonifay, W. E., \& Cai, L. (2017). On the complexity of item response
theory models. \emph{Multivariate Behavioral Research}, \emph{52}(4),
465--484. \url{http://doi.org/10.1080/00273171.2017.1309262}

\hypertarget{ref-Bonifay:2017b}{}
Bonifay, W. E., Lane, S. P., \& Reise, S. P. (2017). Three concerns with
applying a bifactor model as a structure of psychopathology.
\emph{Clinical Psychological Science}, \emph{5}, 184--186.

\hypertarget{ref-Burnham:2002}{}
Burnham, K. P., \& Anderson, D. R. (2002). \emph{Model selection and
multimodel inference: A practical information-theoretic approach} (2nd
ed.). New York: Springer.

\hypertarget{ref-Cliff:1996}{}
Cliff, N. (1996). Answering ordinal questions with ordinal data using
ordinal statistics. \emph{Multivariate Behavioral Research},
\emph{31}(3), 331--350. \url{http://doi.org/10.1207/s15327906mbr3103_4}

\hypertarget{ref-Cohen:1988}{}
Cohen, J. (1988). \emph{Statistical power analysis for the behavioral
sciences} (2nd ed.). Hillsdale, NJ: Lawrence Erlbaum Associates.

\hypertarget{ref-Cudeck:1983}{}
Cudeck, R., \& Browne, M. W. (1983). Cross-validation of covariance
structures. \emph{Multivariate Behavioral Research}, \emph{18},
147--167.

\hypertarget{ref-Donnellan:2016}{}
Donnellan, M. B., Ackerman, R. A., \& Brecheen, C. (2016). Extending
structural analyses of the Rosenberg self-esteem scale to consider
criterion-related validity: Can composite self-esteem scores be good
enough? \emph{Journal of Personality Assessment}, \emph{98}, 169--177.

\hypertarget{ref-Grace:2008}{}
Grace, J. B., \& Bollen, K. A. (2008). Representing general theoretical
concepts in structural equation models: The role of composite variables.
\emph{Evironmental and Ecological Statistics}, \emph{15}, 191--213.
\url{http://doi.org/10.1007/s10651-007-0047-7}

\hypertarget{ref-Hansen:2001}{}
Hansen, M. H., \& Yu, B. (2001). Model selection and the principle of
minimum description length. \emph{Journal of the American Statistical
Association}, \emph{96}, 746--774.
\url{http://doi.org/10.1198/016214501753168398}

\hypertarget{ref-Joe2006}{}
Joe, H. (2006). Generating random correlation matrices based on partial
correlations. \emph{Journal of Multivariate Analysis}, \emph{97}(10),
2177--2189. \url{http://doi.org/10.1016/j.jmva.2005.05.010}

\hypertarget{ref-semTools}{}
Jorgensen, T. D., Pornprasertmanit, S., Schoemann, A. M., \& Rosseel, Y.
(2019). \emph{\texttt{semTools}: Useful tools for structural equation
modeling}. Retrieved from
\url{https://CRAN.R-project.org/package=semTools}

\hypertarget{ref-Lai2017}{}
Lai, K., Green, S. B., \& Levy, R. (2017). Graphical displays for
understanding SEM model similarity. \emph{Structural Equation Modeling:
A Multidisciplinary Journal}, \emph{24}(6), 803--818.
\url{http://doi.org/10.1080/10705511.2017.1334206}

\hypertarget{ref-Lewandowski2009}{}
Lewandowski, D., Kurowicka, D., \& Joe, H. (2009). Generating random
correlation matrices based on vines and extended onion method.
\emph{Journal of Multivariate Analysis}, \emph{100}(9), 1989--2001.
\url{http://doi.org/10.1016/j.jmva.2009.04.008}

\hypertarget{ref-Marsh1994}{}
Marsh, H. W., \& Balla, J. (1994). Goodness of fit in confirmatory
factor analysis: The effects of sample size and model parsimony.
\emph{Quality and Quantity}, \emph{28}(2), 185--217.
\url{http://doi.org/10.1007/BF01102761}

\hypertarget{ref-Mplus:2010:B}{}
Muthèn, L. K., \& Muthèn, B. O. (1998--2010). \emph{Mplus user's guide}
(Sixth). Los Angeles, CA: Muthèn \& Muthèn.

\hypertarget{ref-muthukrishna2019problem}{}
Muthukrishna, M., \& Henrich, J. (2019). A problem in theory.
\emph{Nature Human Behaviour}, \emph{3}(3), 221--229.

\hypertarget{ref-Myung2005:IC}{}
Myung, I. J., Pitt, M. A., \& Kim, W. (2005). Model evaluation, testing
and selection. In K. Lamberts \& R. Goldstone (Eds.), \emph{Handbook of
cognition} (pp. 422--436). Thousand Oaks, CA: Sage.

\hypertarget{ref-OpenMx}{}
Neale, M. C., Hunter, M. D., Pritikin, J. N., Zahery, M., Brick, T. R.,
Kickpatrick, R. M., \ldots{} Boker, S. M. (2016). OpenMx 2.0: Extended
structural equation and statistical modeling. \emph{Psychometrika},
\emph{81}, 535--549.

\hypertarget{ref-Neff2017}{}
Neff, K. D., Whittaker, T. A., \& Karl, A. (2017). Examining the factor
structure of the self-compassion scale in four distinct populations: Is
the use of a total scale score justified? \emph{Journal of Personality
Assessment}, \emph{99}(6), 596--607.
\url{http://doi.org/10.1080/00223891.2016.1269334}

\hypertarget{ref-RColorBrewer}{}
Neuwirth, E. (2014). \emph{RColorBrewer: ColorBrewer palettes}.
Retrieved from \url{https://CRAN.R-project.org/package=RColorBrewer}

\hypertarget{ref-Pearl:1995:IC}{}
Pearl, J., \& Verma, T. S. (1995). A theory of inferred causation. In D.
Prawitz, B. Skyrms, \& D. Westerstahl (Eds.), \emph{Logic, methodology
and philosophy of science ix} (Vol. 134, pp. 789--811). Elsevier.
\url{http://doi.org/https://doi.org/10.1016/S0049-237X(06)80074-1}

\hypertarget{ref-Pitt2002}{}
Pitt, M. A., Myung, I. J., \& Zhang, S. (2002). Toward a method of
selecting among computational models of cognition. \emph{Psychological
Review}, \emph{109}(3), 472--491.
\url{http://doi.org/10.1037/0033-295x.109.3.472}

\hypertarget{ref-Preacher:PhD}{}
Preacher, K. J. (2003). \emph{The role of model complexity in the
evaluation of structural equation models} (PhD thesis). The Ohio State
University.

\hypertarget{ref-Preacher2006}{}
Preacher, K. J. (2006). Quantifying parsimony in structural equation
modeling. \emph{Multivariate Behavioral Research}, \emph{41}(3),
227--259. \url{http://doi.org/10.1207/s15327906mbr4103_1}

\hypertarget{ref-clusterGeneration}{}
Qiu, W., \& Joe., H. (2015). \emph{ClusterGeneration: Random cluster
generation (with specified degree of separation)}. Retrieved from
\url{https://CRAN.R-project.org/package=clusterGeneration}

\hypertarget{ref-Reise2013}{}
Reise, S. P., Bonifay, W. E., \& Haviland, M. G. (2013). Scoring and
modeling psychological measures in the presence of multidimensionality.
\emph{Journal of Personality Assessment}, \emph{95}(2), 129--140.
\url{http://doi.org/10.1080/00223891.2012.725437}

\hypertarget{ref-Reise:2016}{}
Reise, S. P., Kim, D. S., Manslof, M., \& Widaman, K. F. (2016). Is the
bifactor model a better model or is it just better at modeling
implausible responses? Application of iteratively reweighted least
squares to the Rosenberg self-esteem scale. \emph{Multivariate
Behavioral Research}, \emph{51}, 818--838.

\hypertarget{ref-Reise2016}{}
Reise, S. P., Kim, D. S., Mansolf, M., \& Widaman, K. F. (2016). Is the
bifactor model a better model or is it just better at modeling
implausible responses? Application of iteratively reweighted least
squares to the rosenberg self-esteem scale. \emph{Multivariate
Behavioral Research}. \url{http://doi.org/10.1080/00273171.2016.1243461}

\hypertarget{ref-Reise:2007}{}
Reise, S. P., Morizot, J., \& Hays, R. D. (2007). The role of the
bifactor model in resolving dimensionality issues in health outcomes
measures. \emph{Quality of Life Research}, \emph{16}, 19--31.

\hypertarget{ref-Rissanen:1978}{}
Rissanen, J. (1978). Modeling by shortest data description. \emph{Automatica}, \emph{14}, 465--471. \url{http://doi.org/10.1016/0005-1098(78)90005-5}

\hypertarget{ref-Rosenberg:1965}{}
Rosenberg, M. (1965). \emph{Society and the adolescent self-image}.
Princeton, NJ: Princeton University Press.

\hypertarget{ref-lavaan}{}
Rosseel, Y. (2012). lavaan: An R package for structural equation
modeling. \emph{Journal of Statistical Software}, \emph{48}(2), 1--36.
Retrieved from \url{http://www.jstatsoft.org/v48/i02/}

\hypertarget{ref-Savalei2014a}{}
Savalei, V., \& Falk, C. F. (2014). Recovering substantive factor
loadings in the presence of acquiescence bias: A comparison of three
approaches. \emph{Multivariate Behavioral Research}, \emph{49},
407--424.

\hypertarget{ref-Skrondal:2004:B}{}
Skrondal, A., \& Rabe-Hesketh, S. (2004). \emph{Generalized latent
variable modeling: Multilevel, longitudinal, and structural equation
models}. Boca Raton, FL: Chapman; Hall/CRC.

\hypertarget{ref-Steiger1980}{}
Steiger, J. H., \& Lind, J. C. (1980, May). Statistically based tests
for the number of common factors. Paper presented at the annual Spring
Meeting of the Psychometric Society, Iowa City, IA.

\hypertarget{ref-Tucker1973}{}
Tucker, L. R., \& Lewis, C. (1973). A reliability coefficient for
maximum likelihood factor analysis. \emph{Psychometrika}, \emph{38}(1),
1--10. \url{http://doi.org/10.1007/BF02291170}

\hypertarget{ref-Wilcox:2012}{}
Wilcox, R. R. (2012). \emph{Introduction to robust estimation and
hypothesis testing}. Boston: Academic Press.

\setlength\parindent{24pt} \setlength{\leftskip}{0pt}

\appendix

\section*{Appendix: Generation of Random Correlation Matrices}

The original FORTRAN code for the MCMC algorithm was provided by
Kristopher Preacher (2003) and ported to R and modified slightly by the
authors. Let \(\bfr_k\) be a \(p(p-1)/2\) vector of correlations at
iteration \(k\), where \(p\) is the number of observed variables. In
brief, this approach begins the MCMC chain with the correlation matrix
set to an identity matrix, (i.e., \(\bfr_0 = \bf0\)).\footnote{Or a
  matrix with .5 correlations if only positive correlations are desired,
  as in the original code.} Candidate draws, are computed by
\(\bfr_{k+1} = \bfr_{k} + \gamma \bfz\), where \(\gamma\) is a step
size, and \(\bfz = t^{1/p} \frac{\bfx}{\sqrt{\bfx' \bfx}}\), with
\(\bfx\) and \(t\) randomly drawn at each iteration from an independent
normal distribution, \(\bfx \sim \mathcal{N}_p(\bf0,\bfI)\), and uniform
distribution, \(t \sim \text{unif}(0,1)\). Candidate draws are only
rejected if they result in a non-positive-definite matrix, or if
correlation values exceed allowable values (i.e., within \(\pm 1\)). In
order to reduce these possibilities with large correlation matrices,
smaller step sizes are required, which in turn then requires more
iterations between draws to reduce autocorrelations. That is, as in
typical MCMC applications a number determines \emph{thinning} - or the
number of iterations between saving random correlation matrices - and
only a subset of iterations (e.g., \(n=10,000\) out of \(200,000\)) are
saved. The step size (from .56 to .1) and number of iterations (from
200,000 to 10 million) are pre-set under the original algorithm for a
range of observed variables from 3 to approximately 16.

Our modifications of the code included an increase to a default of 5
million iterations for the MCMC algorithm. Parallel processing can be
used by creating \emph{m} independent chains for generating correlation
matrices. The total number of iterations is held constant where possible
by dividing the total number of iterations and draws equally among the
\emph{m} chains. In the case of many processing cores, this could lead
to very few iterations per chain. To prevent this, a minimum number of
iterations per chain is set at 10,000. In both cases, these options are
modifiable by passing a list to an additional argument,
\texttt{mcmc.args}, and documentation on possible options is provided in
the \emph{ockhamSEM} package.

Lewandowski et al. (2009) introduced the ability to generate correlation
matrices with the \emph{vine} and \emph{onion} methods, which are faster
than the MCMC algorithm. The MCMC algorithm may still generate many
correlation matrices that must be discarded due to lack of positive
definiteness. The vine method is based on work by Joe (2006), in which
partial correlations are generated from a linearly transformed Beta
distribution and transformed into product moment correlations. The
computations involved can be illustrated using \emph{C-vines}, which
define the dependency structure among the variables using a graphical
model. We do not pursue the vine method further due to the need to
further study involved tuning parameters that may affect the space for
the randomly generated matrices. The method we pursue in the current
paper is the onion method which allows uniform sampling ``over the space
of correlation matrices'' (p.~1998). The onion method constructs random
correlation matrices recursively, starting with a single dimension and
adding additional dimensions in later steps. Lewandowski et al. (2009)
provide a detailed description of the method as it relates to elliptical
distributions. Either approach is computationally fast, and these
authors report generation of many (5,000) large correlation matrices
(e.g., \(80\times 80\)), in only a few seconds (using compiled C code)
or a minute or two (using Matlab). In the present application, we use
the \emph{clusterGeneration} package (Qiu \& Joe, 2015). As neither
method requires iterations as does MCMC, parallel processing can be
easily conducted by dividing the number of generated correlation
matrices among \emph{m} processing cores.

While we expect that most applications using the onion method will use
correlation matrices generated as-is, we have also implemented the
following ad-hoc procedure for restricting correlation matrix generation
to the space of all positive correlations. Let \(\bfR\) be a generated
correlation matrix. To ensure all positive correlations, we compute a
new correlation matrix by \(\tilde{\bfR} = .5(\bfR + \bf1)\), where
\(\bf1\) is a matrix of 1's of conformable dimensions. We offer no proof
at the moment that this results in uniform sampling from this data
space, however, we note that this approach in many cases results in
similar conclusions regarding fit propensity as the MCMC algorithm.

\end{document}